\documentstyle[12pt,epsfig]{article}

\newcommand{\lsim}{\mathrel{\lower4pt\hbox{$\sim$}}
\hskip-12.5pt\raise1.6pt\hbox{$<$}\;}

\newcommand{\gsim}{\mathrel{\lower4pt\hbox{$\sim$}}
\hskip-12.5pt\raise1.6pt\hbox{$>$}\;}

\begin{document}

\noindent \hspace*{10cm}UCRHEP-T254

\begin{center}
{\bf Flavor changing single top quark 
production channels at $e^+e^-$ colliders 
in the effective Lagrangian description}

\vspace{.3in}

S. Bar-Shalom\footnote{email: shaouly@phyun0.ucr.edu} and J. Wudka\footnote{email: jose.wudka@ucr.edu} \\
\end{center}
\vspace{.1in}

\noindent
Department of Physics, University of California, Riverside CA 92521.\\
\vspace{0.1in}

\begin{center}
{\bf Abstract}\\
\end{center}

We perform a global analysis of the sensitivity of LEP2 and $e^+e^-$ 
colliders with a c.m. energy in the range 500 - 2000 GeV to new 
flavor-changing single top quark production 
in the effective Lagrangian approach. 
The processes considered are sensitive to 
new flavor-changing effective vertices 
such as $Ztc$, $htc$, four-Fermi 
$tcee$ contact terms as well as a right-handed $Wtb$ coupling.
We show that $ e^+ e^- $ colliders are most sensitive to the 
physics responsible for the contact $tcee$ vertices.
For example, it is found that the recent data from the 189 GeV LEP2 run 
can be used to 
rule out any new flavor physics that can generate these four-Fermi 
operators up to energy scales of $\Lambda \gsim 0.7 - 1.4$ TeV, 
depending on the type of the four-Fermi interaction. 
We also show that a corresponding limit of $\Lambda \gsim 1.3 - 2.5$ and 
$\Lambda \gsim 17 - 27$ TeV can be reached at the future 200 GeV LEP2 run and 
a 1000 GeV  $e^+e^-$ collider, respectively.
We note that these limits are much stronger than the typical limits 
which can be placed on flavor diagonal four-Fermi couplings. 
Similar results hold for $\mu^+\mu^-$ colliders and for 
$t \bar u$ associated production. 
Finally we briefly
comment on the necessity of measuring {\em all} flavor-changing effective
vertices as they can be produced by different types of heavy
physics.
\pagebreak

\section{Introduction}

One of the fundamental unresolved issues in high-energy physics is the
origin of the observed (quark) flavor structure. 
Within the Standard Model (SM) flavor-changing processes
are controlled by the scalar sector, and are such that tree-level
Flavor-Changing  Neutral Currents (FCNC) are absent.
This opens the possibility of
using the corresponding flavor-changing processes to probe new physics
whose effects may include appreciable violation of natural flavor conservation 
already at energies probed by present high energy colliders.
For this reason, searching for new flavor-changing dynamics 
will be one of the major goals of the next generation
of high energy colliders such as an $e^+e^-$ Next Linear Collider (NLC) 
\cite{nlcreviews}. 

The top quark, which is the least tested fermion in the SM,
can play an important role in our understanding of flavor dynamics since
its large mass makes it more sensitive to certain types of flavor
changing interactions. 
In particular, $t \to c$ (or $t \to u$) transitions 
which may lead to FCNC signals in high energy colliders, 
offer a unique place for testing the SM flavor structure.
Below we note that,
in addition to direct observations in top production and decays, the
gauge structure of the SM can be used to constrain flavor-changing
processes involving the top-quark through existing data on $B$ meson
decays. 

Top-charm flavor-changing processes can be studied either in 
$t \to c$ decays or in $t \bar c$ pair production in collider experiments.
In the SM such decays \cite{ttocsm1,ttocsm2} 
and production 
\cite{tcprodsm} processes are unobservably small since they
occur at the one-loop level and in addition are GIM suppressed.
Thus, any signal of such $t-c$ transitions will be a clear evidence 
of new flavor physics beyond the SM.
This fact has led to a lot of theoretical activity involving 
top-charm transitions within some specific popular models beyond the SM. 
For example, 
studies of $t \to c$ decays in Multi Higgs doublets 
models (MHDM) \cite{ttocsm1,ttocmhdm1,ttocmhdm2,ttocmhdm3}, 
in supersymmetry with 
R-parity conservation \cite{ttocsusy} and with R-Parity 
violation \cite{ttocrp1,ttocrp2}, and studies of $t \bar c$ production in 
MHDM \cite{ttocmhdm2,ttocmhdm3,tcprodmhdm}, in supersymmetry with 
R-Parity violation \cite{ttocrp2,tcprodrp} and in models with 
extra vector-like quarks \cite{prl82p1628}.
In this paper we will use instead a model independent approach 
\cite{b.and.w} to investigate $t \bar c$ (and\footnote{Throughout 
this paper we will loosely refer to a $t \bar c + \bar t c$ final state
by $t \bar c$. The contributions from the
charged conjugate $\bar t c$ state are included in our numerical results 
unless explicitly stated otherwise.} $\bar t c$)
pair production in $e^+e^-$ 
colliders such as LEP2 and a NLC 
with c.m. energies of 500 - 2000 GeV \cite{nlcreviews}. 

It is important to stress the advantage of studying $t \bar c$ production 
over $ t \to c$ decays signals in high energy collider experiments in such 
a model independent approach. While $t \to c$ decays will be
suppressed by powers of $m_t/\Lambda$, where $\Lambda$ indicates 
the heavy physics energy scale, the corresponding suppression 
factor for $t \bar c$ production processes are proportional to a power
of $ E_{CM}/\Lambda$, where $E_{CM}$ is the c.m. energy of the collider.
From the experimental point of view, 
a $t \bar c$ signal has some very distinct characteristics, in particular,
it has the unique signature of producing a single $b$-jet in the final state. 
In a recent paper \cite{bparity} 
we have observed that the SM cross sections for processes with
an odd number of $b$-jets in the final state are extremely small, which
allows the definition of a new 
approximately conserved quantum number: $b$-Parity ($b_P$). Processes
with even or odd number of $b$ jets have $b_P=1$ and $b_P=-1$ respectively.
Thus the $b_P$-odd process $e^+e^- \to t \bar c$ 
can be detected using the simple $b$-jet counting method suggested 
in \cite{bparity}, and is essentially free of any
SM irreducible background.\footnote{There is, of course, a reducible
background generated due to reduced $b$-tagging efficiency, 
see \cite{bparity}.} 

Several model-independent studies of $t \bar c$ pair production have
appeared in the literature, where the signatures and observability of
these flavor violating processes were investigated in
$e^+e^-$ colliders \cite{prl82p1628,eetcindep1,joan,hep9712394}, 
hadronic colliders \cite{davidsher} and 
$\gamma \gamma$ colliders \cite{gamgamtc}. 
The present paper extends the results obtained in \cite{eetcindep1,joan,hep9712394}
by performing a model-independent analysis in a wider variety of channels. 
In particular, we explore the sensitivity of 
$e^+e^-$ colliders to all relevant effective operators that 
can give rise to $t \bar c$ production in $e^+e^-$ colliders with a 
c.m. energy ranging from 189 GeV (LEP2) to 2000 
GeV.\footnote{To be specific we 
consider reactions in $e^+e^-$ colliders, but the analysis performed is 
clearly extendable to muon colliders.}
We consider the $2 \to 2$
processes $e^+e^- \to t \bar c$, $e^+ e^- \to Zh$ followed 
by $h \to t \bar c$, where $h$ is the SM Higgs-boson, and 
the $t$-channel fusion processes $W^+W^-,~ZZ \to t \bar c$. 
These reactions can proceed via new $Ztc$, $htc$ and $Wtb$ couplings as well 
as through new $tcee$ four-Fermi effective operators that have not been
previously considered in this context.

We argue that since the effective interactions are but the low energy
manifestations of an underlying theory, and assuming this heavy theory is a
gauge theory containing fermions, scalars 
and gauge-bosons, then some of the effective vertices 
that contribute to 
$e^+e^- \to t \bar c$ are expected to be 
suppressed and will produce very small effects
(the $Ztc$ and $\gamma tc$ magnetic-type couplings
considered in \cite{joan,hep9712394} fall into this category). 
We therefore do not include such couplings 
(see section 2). We will concentrate
on those vertices for which general principles do not mandate a small
coefficient.

Following the above viewpoint, our study indicates that
the reaction $ e^+e^- \to t \bar c$ is most sensitive to effective
four-Fermi flavor-changing interactions. It is found, for example, that if 
the coupling strength of the four-Fermi interactions is 
of ${\cal O} (1/\Lambda^2)$ as expected by naturalness, 
then tens to hundreds $t \bar c$ events should 
show up already at LEP2 energies when $ \Lambda \lsim 1 $ TeV.
Alternatively, if no $e^+ e^- \to t \bar c$ signal is observed, 
then the limits that can be placed on the energy scale $\Lambda$ of 
such four-Fermi effective operators are quite strong; 
the data accumulated at the recent 189 GeV LEP2 run can already place 
the limit $\Lambda \gsim 0.7 -1.4$ TeV, while 
$\Lambda \gsim 1.5 -2.5$ TeV will be achievable at a 200 GeV LEP2 and
reaching 
$\Lambda \gsim 17-27$ TeV at a NLC with a c.m. energy of 1000 GeV 
(depending on the type of the four-Fermi operator).
It is remarkable that a 500 - 1000 GeV $e^+e^-$ collider can place a bound 
on such four-Fermi dynamics which is almost 20--30 times larger than
its c.m. energy. These limits can be compared, for example, 
with the bound $\Lambda \gsim 5$ TeV 
that can be obtained on $eeee$ and $ttee$ four-Fermi operators 
by studying the reactions $e^+e^- \to e^+e^-$ \cite{jose2}
and $e^+e^- \to t \bar t$ \cite{grzad} at a NLC;
note, however, that the scales responsible for the $tcee$ and $ttee$ 
(or $eeee$)
vertices need not be the same.
Similarly the effective $tcee$ and $Ztc$ vertices may be produced 
by different physics, e.g., a heavy neutral vector boson 
(for $tcee$) vs. heavy vector-like quarks (for $Ztc$) \cite{prl82p1628};
there are new physics possibilities which are best probed through
$Ztc$ interactions.
In all cases the sensitivity to $
\Lambda$ will be significantly degraded if the couplings are $ \ll 1$.

Flavor violating
$Z$ and Higgs ($h$) interactions, such as a possible effective $Ztc$ 
and $htc$ vertices, are
probed via $WW$-fusion processes $ e^+e^- \to W^+W^- \nu_e \bar\nu_e \to 
t \bar c \nu_e \bar\nu_e$, and the Bjorken process 
$e^+e^- \to Zh$ followed by $h \to t \bar c$ for $htc$. For example, if no
$t \bar c \nu_e \bar\nu_e$ signal is observed at 1500 GeV 
(500 GeV) NLC, 
then the limit $\Lambda \gsim 2 $ TeV ($ \gsim 800$ GeV), for a 
SM Higgs mass of $ 250 $ GeV, and assuming 
that the the $htc$ vertices have a coupling strength of 
${\cal O} (v^2/\Lambda^2)$ ($v$ is the vacuum expectation value 
of the SM scalar field).
The effects of new $Ztc$ and $htc$ effective 
couplings on the $ZZ$-fusion process 
$e^+e^- \to ZZ e^+e^- \to t \bar c e^+e^-$ are too small to be detected 
at a NLC. 
The same is true of a right-handed $Wtb$ coupling
in the reaction $e^+e^- \to W^+W^- \nu_e \bar\nu_e \to 
t \bar c \nu_e \bar\nu_e$, even when assuming a coupling with a strength of 
${\cal O} (v^2/\Lambda^2)$, the bound allowed by naturality.

We note that, since charm quark mass effects are negligible at 
high energy $e^+e^-$ colliders, our results equivalently apply 
to $t \bar u$ pair production. In particular, to effective operators 
generating the corresponding $tu$ flavor-changing interactions.     
  
The paper is organized as follows: in section 2 we describe the 
effective Lagrangian framework and extract the Feynman rules for 
the new effective vertices. In section 3 we discuss the effects 
of new $Ztc$ vector couplings and $tcee$ four-Fermi interactions 
in $e^+e^- \to t \bar c$ and $W^+W^-,~ZZ \to t \bar c$.
In section 4 we consider the contribution of new 
$htc$ scalar couplings to $e^+e^- \to Zh \to Zt \bar c$ and 
to $W^+W^- \to t \bar c$. In section 5 we investigate the 
effects of a new right-handed $Wtb$ coupling on the process
$W^+W^- \to t \bar c$ and in section 6 we summarize our results.

\section{The effective Lagrangian description and \boldmath{$t \bar c$} 
production at \boldmath{$e^+e^-$} colliders} 

There are two different theoretical paths one can adopt to investigate 
physics beyond the SM. In the first, one uses a
specific model to calculate such effects. The second is to follow a
model-independent approach where the effects of any given high energy model
is parameterized by the coefficients of a series of
effective operators without reference to any specific underlying 
theory. The power of the 
model-independent approach lies in its generality, its potential
deficiency is the large number of constants which might {\it a
priori} contribute to any given reaction. In this paper we follow the
second route.
 
Our basic assumption will be that there is a gauge theory underlying the
SM, whose scale $ \Lambda $ is well
separated from the Fermi scale. Under these circumstances the low energy
limit of the theory will consist of the SM Lagrangian plus
corrections represented by a series of effective operators $ {\cal O}_i $
constructed using the SM fields and whose
coefficients are suppressed by powers of $ 1/ \Lambda $
\begin{eqnarray}
{\cal L}_{eff} = {\cal L}_{SM} + \sum_{n=5}^\infty
\frac{1}{\Lambda^{n-4}}\sum_i \alpha_i {\cal O}_i^n~,\label{ltot}
\end{eqnarray}
where each $ {\cal O } $ respects the gauge symmetries of the SM but not
necessarily its global symmetries.\footnote{For example
operators of dimension 5, if present,
necessarily violate lepton number, \cite{b.and.w}.} 
The dominating effects
are usually generated by the lowest-dimensional operators contributing
to the process of interest (there are, however, some exceptions, see \cite{b.and.w}).
For the flavor-violating processes considered here the only relevant
operators are those of dimension 6, if these are absent there will be no 
observable signal.

In the following discussion we will assume, for definiteness, that the
theory underlying the SM is weakly coupled; but we expect our
results to hold in general. The reason is that both in  weakly and
strongly coupled (natural) theories, the dominating
flavor-changing effects (at least for the processes considered)
are produced by the four-Fermi contact interactions, for which
naturality allows the largest coefficients \cite{naturality}.

Now, it is important to note that general considerations
require certain bounds for the coefficients 
$\alpha_i$ in (\ref{ltot}).
For weakly-coupled underlying theories the key point is that
the effective operators 
may correspond to either tree-level 
or loop exchanges of 
the heavy fields. Loop-generated interactions are 
suppressed by factors of $\sim 1/16\pi^2$ (and by
powers of the coupling constants) compared to 
the tree-level induced operators. 
One therefore expects the effects 
of the high energy theory to manifest themselves
predominantly through tree-level generated (TLG) operators. 
In what follows we consider only TLG
operators and neglect those generated by loops involving the heavy particles. 

The observables studied in this
paper cannot distinguish between models with large values of $ \Lambda$
having tree-level flavor-changing interactions and those models
with lower values of $\Lambda$ for which flavor-changing processes
occur only via loops. But this ambiguity is only academic when
discussing heavy physics virtual effects, 
as neither of these situations will produce measurable effects.
Only models whose scales lie below $ \sim 10 $ TeV
and which generate flavor violation at tree-level will be observed through
the processes considered in this paper.

We stress that this approach is in general different 
from the one adopted in many
previous investigations which use the effective Lagrangian description 
to study new physics in present and future colliders. 
For example, we do not include anomalous 
dipole-like operators of the form ($V=\gamma$ or $Z$)
\begin{eqnarray}
i e \bar t \frac{\sigma_{\mu \nu} q^\nu}{m_t} 
(\kappa_V - i \tilde\kappa_V \gamma_5) c
V^\mu \label{dipolelike}~,
\end{eqnarray}
in the reaction $e^+e^- \to t \bar c$, 
since the coefficients from these vertices
are much smaller than those of the $tcee$ four-Fermi
vertices. In fact, assuming the 
physics underlying the SM is weakly-coupled, the typical
size of the coefficients are $\kappa_V,~\tilde\kappa_V 
\sim ( v^2/\Lambda^2) \times 1/16 \pi^2 \sim 4 \times 10^{-4}$, for 
$\Lambda \sim 1$ TeV. Thus the corresponding contributions are
subdominant despite their rapid growth with energy. 
If instead $\kappa_V$ or 
$\tilde\kappa_V$ $\sim {\cal O}(1)$ ($\sim {\cal O}(0.1)$) is used - 
as required 
in order to have an appreciable $t \bar c$ production rate - 
what in fact is being done is to assume that the scale of
``new physics'' is $\Lambda \sim v/4 \pi \sim 20$ GeV 
($\Lambda \sim v/4 \sim 60$ GeV), which is of course 
unacceptable bearing the existing experimental evidence of the 
validity of the SM at these energy scales.
Another loop induced effective operator that can give rise to a $t \bar c$
final state and that falls into this category is a $VVtc$ ($V=W$ or $Z$) 
contact term. In the following we will
neglect these and similar contributions. 

\begin{figure}[htb]
\psfull
 \begin{center}
 \leavevmode
 \epsfig{file=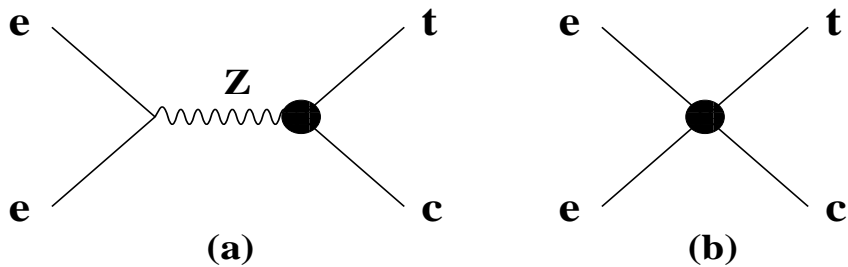,height=6cm,width=6cm,bbllx=0cm,bblly=2cm,bburx=20cm,bbury=25cm,angle=0}
 \end{center}
\caption{\emph{Feynman diagrams that give rise to $e^+e^- \to t \bar c$ 
in the presence of (a) a new $Ztc$ coupling and (b) a new 
$tcee$ four-Fermi coupling. The new effective vertex is denoted by a 
heavy dot.}}
\label{fig1}
\end{figure}

In contrast, new vector and pseudo-vector couplings in the 
$Ztc$ vertex (note that the corresponding $\gamma tc$ 
couplings are forbidden by 
$U(1)$ gauge invariance) as well as new four-Fermi $tcee$ interactions, 
can arise from TLG effective operators
and their coefficients can, therefore, take values typically of the order of 
$\sim (v^2/\Lambda^2) $. If present, these operators will give the 
dominant contribution to $t \bar c$ production; if these interactions are 
either absent or suppressed at tree-level, the $ t \bar c $ production
rate will be unobservably small. In the following we will investigate the
possible effects due to TLG operators
assuming no additional suppression factors are present.

We first list all the TLG effective operators contributing to
$t \bar c$ pair production 
in high energy $e^+e^-$ colliders via the processes

\begin{eqnarray}
&&e^+e^- \to t \bar c \label{tc}~,\\ 
&&e^+e^- \to Zh \to Z t \bar c \label{tcz}~,\\
&&e^+ e^- \to W^+ W^- \nu_e \bar\nu_e \to 
t \bar c \nu_e \bar\nu_e \label{tcnunu}~,\\
&&e^+ e^- \to ZZ e^+ e^- \to t \bar c e^+ e^- \label{tcee}~.
\end{eqnarray}

\begin{figure}[htb]
\psfull
 \begin{center}
 \leavevmode
 \epsfig{file=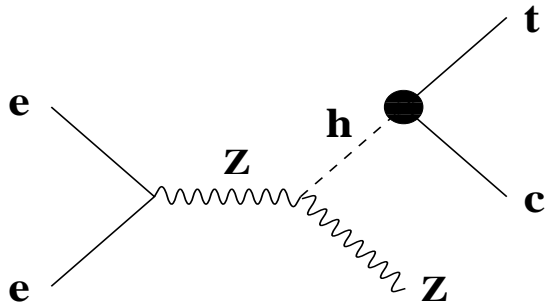,height=6cm,width=6cm,bbllx=0cm,bblly=2cm,bburx=20cm,bbury=25cm,angle=0}
 \end{center}
\caption{\emph{Feynman diagram that give rise to $e^+e^- \to Z t \bar c$
via the Bjorken process $e^+e^- \to Zh$ followed by $h \to t \bar c$, 
in the presence of a new $htc$ coupling. 
The new effective vertex is denoted by a 
heavy dot.}}
\label{fig2}
\end{figure}

\begin{figure}[htb]
\psfull
 \begin{center}
 \leavevmode
 \epsfig{file=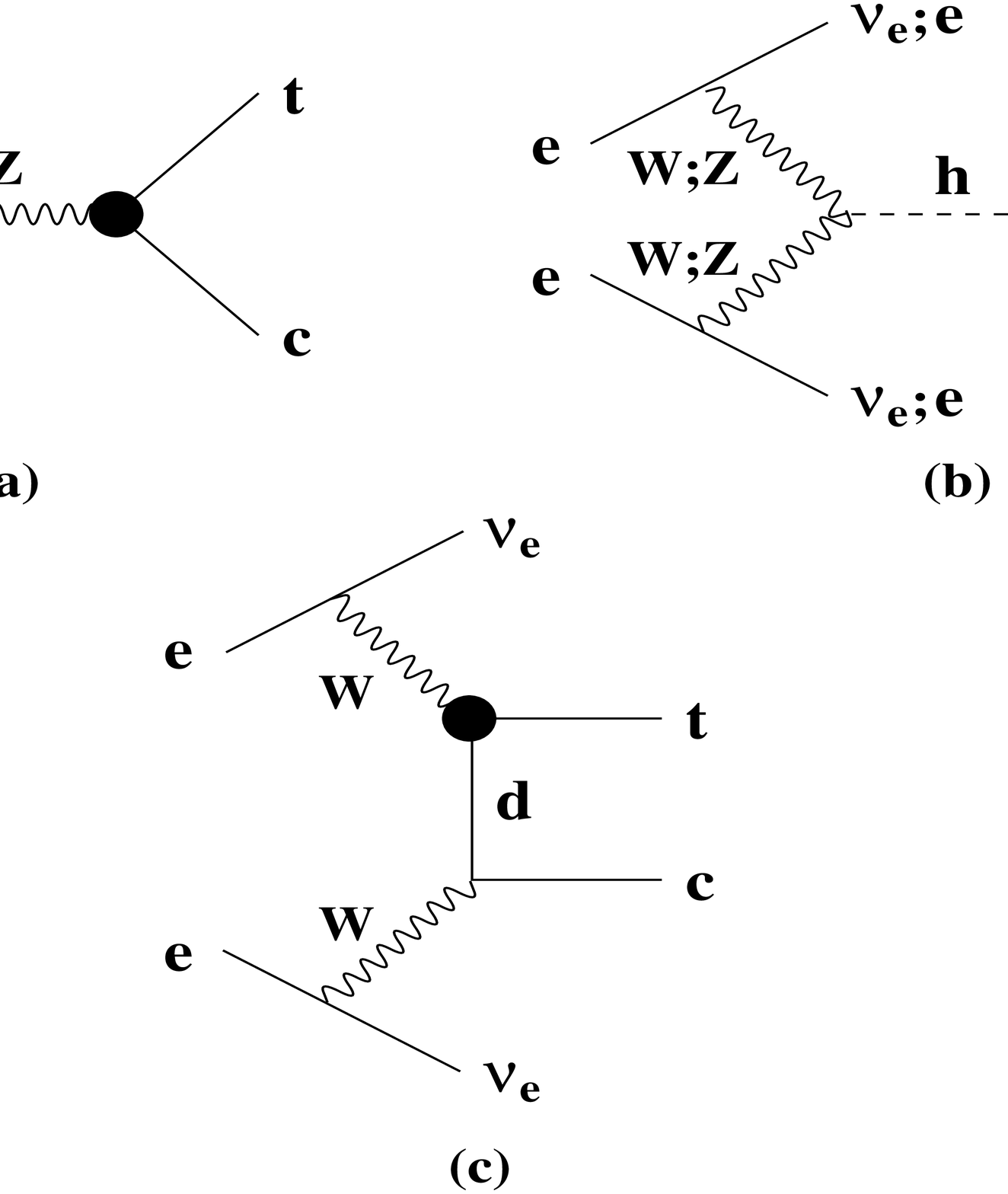,height=8cm,width=6cm,bbllx=0cm,bblly=2cm,bburx=20cm,bbury=25cm,angle=0}
 \end{center}
\caption{\emph{Feynman diagrams that give rise to the 
$WW$-fusion process $e^+e^- \to t \bar c \nu_e \bar\nu_e$, 
in the presence of (a) a new $Ztc$ coupling, (b) a new $htc$ coupling 
and (c) a new $Wtd$ coupling where $d=d,~s$ or a $b$-quark. 
Also plotted in (b) is the Feynman diagram that gives rise to the 
$ZZ$-fusion process 
$e^+e^- \to t \bar c e^+e^- $ in the presence of a new $htc$ coupling. 
The new effective vertex is denoted by a 
heavy dot.}}
\label{fig3}
\end{figure}

\begin{figure}[htb]
\psfull
 \begin{center}
 \leavevmode
 \epsfig{file=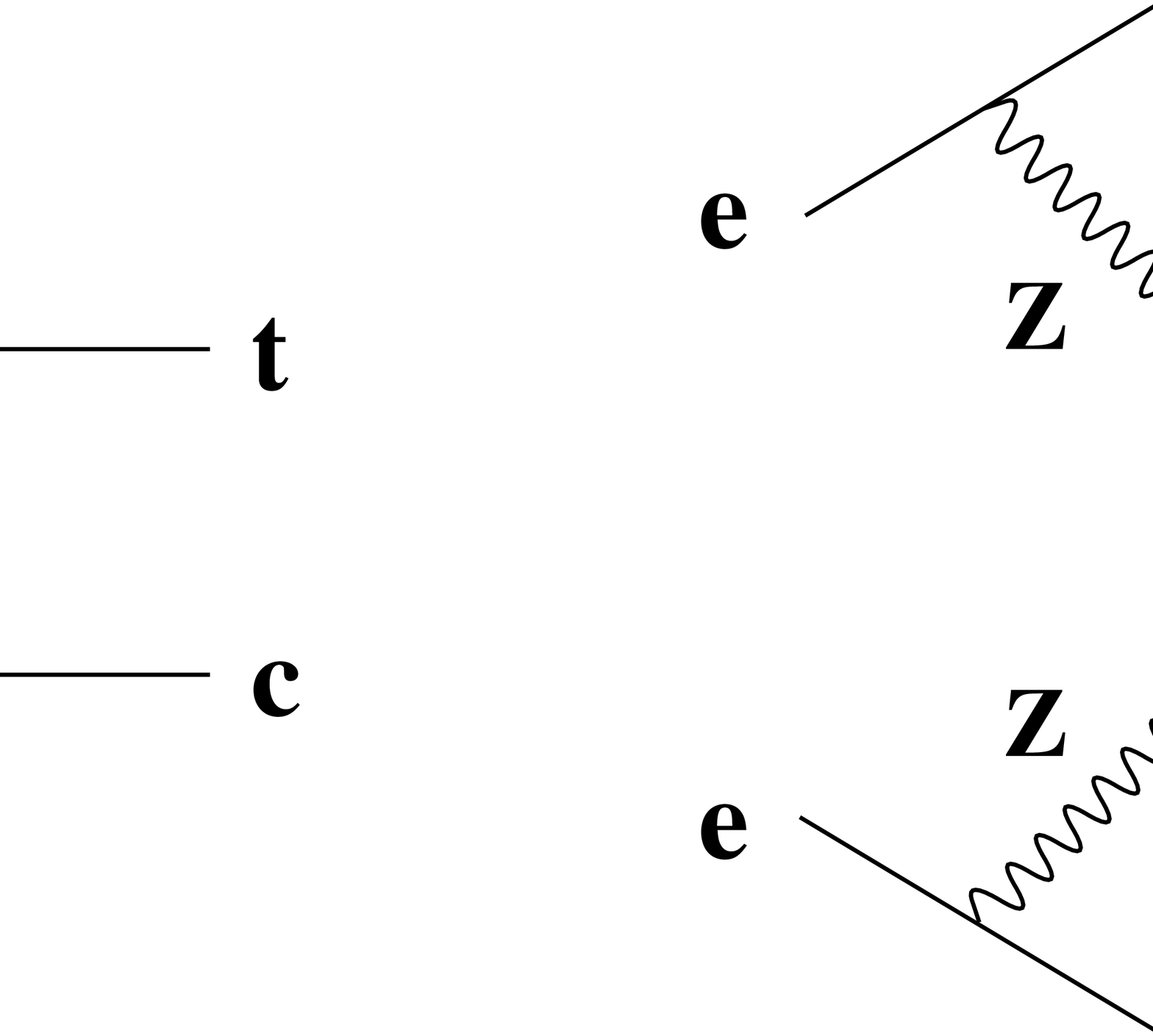,height=6cm,width=6cm,bbllx=0cm,bblly=2cm,bburx=20cm,bbury=25cm,angle=0}
 \end{center}
\caption{\emph{Feynman diagrams that give rise to the 
$ZZ$-fusion process $e^+e^- \to t \bar c e^+e^- $ 
in the presence of a new $Ztc$ coupling. 
The new effective vertex is denoted by a 
heavy dot.}}
\label{fig4}
\end{figure}

\noindent Reaction (\ref{tc}) receives contributions from both an effective 
$Ztc$ interaction (see Fig.~\ref{fig1}(a)) 
and from four-Fermi $tcee$ effective operators 
(see Fig.~\ref{fig1}(b)).
In reaction (\ref{tcz}) we assume real Higgs ($h$) production followed 
by the Higgs decay $h \to t \bar c$, which occurs only in 
the presence of a new $ht c$ interaction as depicted in Fig.~\ref{fig2}. 
Reaction (\ref{tcnunu}) gets contributions from non-standard 
$Zt c$, $htc$ and $W t d$ 
($d$ stands for any of the three down quarks in the SM) vertices 
as depicted in Figs.~\ref{fig3}(a), (b) and (c), respectively. 
Finally, reaction 
(\ref{tcee}) may receive contributions from non-standard 
$ht c$ as well as $Zt c$ vertices 
as shown in Fig.~\ref{fig3}(b) and Figs.~\ref{fig4}(a) and (b).
Below we list the TLG effective operators which give rise to such 
new couplings.

Our notation is the following \cite{b.and.w}: $q$ and $\ell$ denote
left-handed $SU(2)$ quark and lepton doublets, respectively;
$d$, $u$ and $e$ for right-handed ($SU(2)$ singlet) down-quark,
up-quark and charged lepton, 
respectively. The SM 
scalar doublet is denoted by $\phi$ and $D$ is the covariant derivative.
The Pauli matrices are denoted by $ \tau_I $, $I=1,2,3$.
Also, although we suppress generation indices in the effective operators below,
 it should be understood that the quark fields can correspond to different 
flavors in general, i.e., in our case $\bar q$ or $\bar u$ 
is the outgoing top quark and $q$ or $u$ is the incoming charm quark 
(or outgoing anti-charm quark).\\

\subsection{Effective operators generating a $Zt c$ vertex}

There are three tree-level dimension 6 effective operators that can 
generate a new $Z t c$ interaction. These are\footnote{Although 
we do not explicitly include the hermitian conjugate 
operators, it should be clear
that in our case, i.e., $t \bar c$ production,  
the effective operators for 
$\bar t c$ are the hermitian conjugate of those that are given below 
for the $t \bar c$ final state.} \cite{b.and.w}:
\begin{eqnarray}
{\cal O}_{\phi q}^{(1)} &=& i \left( \phi^{\dagger} D_\mu \phi \right ) \left(
\bar q \gamma^\mu q \right) ,\cr
{\cal O}_{\phi q}^{(3)} &=& i \left( \phi^{\dagger} D_\mu \tau^I \phi \right ) 
\left(\bar q \gamma^\mu \tau^I q \right) ,\cr
{\cal O}_{\phi u} &=& i \left( \phi^{\dagger} D_\mu \phi \right ) \left(
\bar d \gamma^\mu d \right) \label{ophiq1q3}~.
\end{eqnarray}

\noindent Writing the new $Zt c$ effective Lagrangian as
\begin{eqnarray}
{\cal L}_{Z_\mu t c} = g { v^2 \over \Lambda^2 } \bar t \gamma_\mu 
\left(a_L^Z L +a_R^Z R \right) c
\label{ztc} ~,
\end{eqnarray}
where $L(R) = (1-(+)\gamma_5)/2$, we can express the left 
and right couplings, $a_L^Z$ and $a_R^Z$, in terms of the corresponding 
coefficients $\alpha_{\phi q}^{(1)},~\alpha_{\phi q}^{(3)}$ and 
$\alpha_{\phi u}$ 
(following our notation in (\ref{ltot})),
\begin{equation}
a_L^Z = \frac{1}{4 c_W} \left( 
\alpha_{\phi q}^{(1)} - \alpha_{\phi q}^{(3)} \right) ~,\qquad
a_R^Z = \frac{1}{4 c_W} \alpha_{\phi u} ~,
\end{equation} 
where $c_W = \cos\theta_W$ and $\theta_W$ is the weak 
mixing angle. 

The operators in (\ref{ophiq1q3}) can be generated at tree-level by heavy
gauge-boson or fermion exchange. 

\subsection{Effective operators generating a new $htc$ vertex}

Apart from the operators in (\ref{ophiq1q3}), 
which give rise also to a new $h t c$ interaction, 
there is an additional operator \cite{b.and.w},
\begin{eqnarray}
{\cal O}_{u \phi} &=& \left( \phi^{\dagger} \phi \right ) \left(
\bar q u \tilde\phi \right) \label{ouphi}~.
\end{eqnarray}
Writing the new $ht c$ interaction Lagrangian as
\begin{eqnarray}
{\cal L}_{h t c} = g {v^2 \over \Lambda^2} \bar t \left(a_L^h L +a_R^h R \right) c 
\label{htc} ~,
\end{eqnarray}
we have (neglecting terms proportional to the charm quark mass)
\begin{equation}
a_L^h = \frac{m_t}{2 g v} \left( 
\alpha_{\phi q}^{(1)} - \alpha_{\phi q}^{(3)} \right)
\label{alh}~,\qquad
a_R^h = \frac{m_t}{2 g v} \left(
\alpha_{\phi u} + { 3 v \over \sqrt{2} \, m_t } \alpha_{u \phi} \right) 
\label{arh}~,
\end{equation}
The heavy excitations which can generate $ {\cal O}_{u \phi } $ at tree-level
are either heavy scalars mixing with the $ \phi $, and/or heavy fermions
mixing with the light fermions and $\phi$. In the first case there is a
contribution only if the mixing occurs through $ {\cal O}(\Lambda )$
cubic couplings and is suppressed in natural theories.

\subsection{Effective operators that generate new $W t d_i$ and $W c d_i$ vertices}

Here there are two operators. One is ${\cal O}_{\phi q}^{(3)}$ in 
(\ref{ophiq1q3}), the second is
\begin{eqnarray}
{\cal O}_{\phi \phi} &=& \left( \phi^{\dagger} \epsilon D_\mu \phi \right ) 
\left(\bar u \gamma^\mu d \right) \label{ouphiphi}~,
\end{eqnarray}
with $\epsilon_{12}=-\epsilon_{21}=1$.

We will parameterize the $W t \bar d_i$ and $W \bar c d_i$ 
($d_i=d,s$ or $b$ for $i=1,2$ or 3, respectively) vertices according to
\begin{eqnarray}
{\cal L}_{W_\mu t \bar d_i} &=& 
{v^2 \over \Lambda^2 } \frac{g}{\sqrt 2} \bar t \gamma_\mu 
\left(V_{3i} L + \delta_{L,i}^t L + \delta_{R,i}^t R \right) d_i \label{wtd} ~,\\
{\cal L}_{W_\mu \bar c d_i} &=& { v^2 \over \Lambda^2 } \frac{g}{\sqrt 2} \bar d_i \gamma_\mu 
\left(V_{2i}^* L + \delta_{L,i}^c L + \delta_{R,i}^c R \right) c 
\label{wcd} ~,
\end{eqnarray}
where $V$ is the CKM matrix.
Thus, if all the relevant coefficients are real 
(as assumed in this paper), then one has
\begin{eqnarray}
\delta_{L,i}^t = \alpha_{\phi q}^{(3)} \mid_i^t ~~,~~
\delta_{R,i}^t = -\frac{1}{2} \alpha_{\phi \phi} \mid_i^t ~~,~~
\delta_{L,i}^c = \alpha_{\phi q}^{(3)} \mid_i^c ~~,~~
\delta_{R,i}^c = -\frac{1}{2} \alpha_{\phi \phi} \mid_i^c . \label{wdef}~
\end{eqnarray}
Notice that, since the operators 
${\cal O}_{\phi \phi}$ and ${\cal O}_{\phi q}^{(3)}$ may have different 
coefficients for different flavors (families) of the up and down quarks;
in order to be as general as possible, 
we have added 
the subscript $i$ and the superscript $t$ or $c$ appropriately.

The heavy excitations that can generate ${\cal O}_{\phi \phi}$ are either a
heavy gauge boson which couples to $ \phi $, or a heavy fermion which couples
to the light fermions and to $ \phi $.

\subsection{Four-Fermi effective operators producing a $tcee$ 
contact interaction}

There are seven relevant four-Fermi operators that contribute to 
$e^+e^- \to t \bar c$
\begin{eqnarray}
{\cal O}_{\ell q}^{(1)} &=& \frac{1}{2} 
\left( \bar\ell \gamma_\mu \ell \right ) 
\left(\bar q \gamma^\mu q \right) \label{4fermi1}~,\\
{\cal O}_{\ell q}^{(3)} &=& \frac{1}{2} 
\left( \bar\ell \gamma_\mu \tau^I \ell \right ) 
\left(\bar q \gamma^\mu \tau^I q \right) \label{4fermi2}~,\\
{\cal O}_{e u} &=& \frac{1}{2} 
\left( \bar e \gamma_\mu e \right ) 
\left(\bar u \gamma^\mu u \right) \label{4fermi3}~,\\
{\cal O}_{\ell q} &=&
\left( \bar\ell e \right ) \epsilon 
\left(\bar q u \right) \label{4fermi4}~,\\
{\cal O}_{q e} &=& 
\left( \bar q e \right ) 
\left(\bar e q \right) \label{4fermi5}~,\\
{\cal O}_{\ell u} &=& 
\left( \bar\ell u \right ) 
\left(\bar u \ell \right) \label{4fermi6}~,\\
{\cal O}_{\ell q^{\prime}} &=& 
\left( \bar\ell u \right ) \epsilon 
\left(\bar q e \right) \label{4fermi7}~.
\end{eqnarray}

\noindent One can also parameterize the most 
general four-Fermi effective Lagrangian 
for the $t \bar c e^+e^- $ interaction in the form
\begin{eqnarray}
{\cal L}_{tcee} = && {1\over\Lambda^2} \sum_{i,j=L,R}
\biggl[ V_{ij} \left({\bar e} \gamma_\mu P_i e \right) 
\left( \bar t \gamma^\mu P_j c \right) 
+ S_{ij} \left( {\bar e} P_i e \right) 
\left( \bar t P_j c \right) \nonumber \\  
&&~~~~~~~~~~~~~+ T_{ij} \left( {\bar e} \sigma_{\mu \nu} P_i e \right) 
\left( \bar t \sigma_{\mu \nu} P_j c \right) \biggr] \label{4fermimatrix}~,
\end{eqnarray}
where $P_{L,R} = (1 \mp \gamma_5)/2$, and 
express these vector-like ($V_{ij}$), scalar-like ($S_{ij}$) 
and tensor-like ($T_{ij}$) couplings in terms of the coefficients 
of the seven four-Fermi operators in (\ref{4fermi1})--(\ref{4fermi7}). 
We get (Fierz-transforming the last four operators)
\begin{eqnarray}
&& 
V_{LL}=\frac{1}{2} \left( \alpha_{\ell q}^{(1)} - 
\alpha_{\ell q}^{(3)} \right)~,~ 
V_{LR}=-\frac{1}{2} \alpha_{\ell u} ~, ~
V_{RR}=\frac{1}{2} \alpha_{e u} ~, ~
V_{RL}=-\frac{1}{2} \alpha_{q e} 
~, \cr
&& 
S_{RR}=-\alpha_{\ell q} + \frac{1}{2} \alpha_{\ell q^{\prime}} ~, ~
S_{LL}=S_{LR}=S_{RL}=0 
~, \cr
&& 
T_{RR}=\frac{1}{8} \alpha_{\ell q^{\prime}}~, ~
T_{LL}=T_{LR}=T_{RL}=0 \label{vst}~.
\end{eqnarray}

\noindent The four-Fermi operators can be generated through the exchange of
heavy vectors and scalars. Note however that the list provided
does not include tensor operators, which have been eliminated
using Fierz transformations. It is therefore possible for a 
tensor exchange to be hidden in a series of operators involving scalars
(and vice-versa). It is noteworthy that no $LL$ tensor or $LL$, $LR$ and 
$RL$ scalar terms are
generated by dimension 6 operators (they can be generated by dimension 8
operators and have coefficients $ \sim ( v/\Lambda)^4$.

\section{\boldmath{$t cee$} four-Fermi interactions and 
\boldmath{$e^+e^- \to t \bar c$}}

As discussed in the previous section, 
there are seven possible TLG four-Fermi effective operators 
(see (\ref{4fermi1})--(\ref{4fermi7})) respecting the SM symmetries;
The effects of such four-Fermi operators have not been investigated 
in $e^+e^- \to t \bar c$;
in this section we calculate the contribution of these operators to
this process.

Using the effective four-Fermi 
Lagrangian piece in (\ref{4fermimatrix}), we obtain 
the amplitude for $e^+e^- \to t \bar c$
\begin{eqnarray}
 && {\cal M}_{tcee} = \frac{1}{\Lambda^2} \sum_{ij} \left\{ 
V_{ij} \left( \bar v_{\bar e} \gamma_\mu P_i u_e \right) 
\left( \bar u_t \gamma^\mu P_j v_c \right) 
+ S_{ij} \left( \bar v_{\bar e} P_i u_e \right) 
\left( \bar u_t P_j v_c \right) \right. \nonumber \\ 
&& \qquad \qquad \qquad\qquad \qquad +\left. 
 T_{ij} \left( \bar v_{\bar e} \sigma_{\mu \nu} P_i u_e \right) 
\left( \bar u_t \sigma_{\mu \nu} P_j v_c \right) \right\} \label{4fermieetc}~,
\end{eqnarray}
where $i,j=L$ or $R$.
Recall that the only non-zero scalar and vector couplings are
$S_{RR}$ and $T_{RR}$.

The cross sections for polarized incoming electrons
and outgoing top quarks (i.e., left or right-handed electron 
and top quark) are then readily calculated 
(recall that we assume all the new couplings to be real)
\begin{eqnarray}
\sigma_{e_L t_L} & = & \sigma(e^-_L e^+ \to t_L \bar c) = 
{\cal C} 
\left[ 2(1+\beta_t) V_{LL}^2 + (1-\beta_t) V_{LR}^2 
\right] \label{sigll} ~,\\
\sigma_{e_L t_R} & = & \sigma(e^-_L e^+ \to t_R \bar c) = 
{\cal C} 
\left[ (1-\beta_t) V_{LL}^2 + 2(1+\beta_t) V_{LR}^2 
\right] \label{siglr} ~,\\ 
\sigma_{e_R t_L} & = & \sigma(e^-_R e^+ \to t_L \bar c) = 
{\cal C} 
\left[ 2(1+\beta_t) V_{RL}^2 + (1-\beta_t) V_{RR}^2 
\right. \nonumber\\
&&\left. ~~~~~~~~~~~~~~~~~~~~~~~ +\frac{1}{2} (1+\beta_t) 
(3 S_{RR}^2 + 16 T_{RR}^2) 
\right] \label{sigrl} ~,\\ 
\sigma_{e_R t_R} & = & \sigma(e^-_R e^+ \to t_R \bar c) = 
{\cal C} 
\left[ (1-\beta_t) V_{RL}^2 + 2(1+\beta_t) V_{RR}^2 
\right. \nonumber\\
&&\left. ~~~~~~~~~~~~~~~~~~~~~~~ +16 (1- \beta_t) T_{RR}^2) 
\right] \label{sigrr} ~,
\end{eqnarray}
%
%
where 
\begin{eqnarray}
{\cal C}=\frac{s}{\Lambda^4} \frac{\beta_t^2}{4 \pi (1+\beta_t)^3} 
\label{const}~,
\end{eqnarray}
and $\beta_t=(s-m_t^2)/(s+m_t^2)$. 
The total unpolarized cross section for production of 
$t \bar c + \bar t c$ pairs is then 
\begin{eqnarray}
\sigma_{tc} = \sigma(e^- e^+ \to t \bar c + \bar t c) = 
\sum_{i,j=L,R} \sigma_{e_i t_j}~.
\end{eqnarray} 

Notice that, by assumption, 
such four-Fermi interactions are induced by 
exchanges of a heavy field in the underlying high energy theory for which
one is replacing the heavy particle propagator by $1/\Lambda^2$.
Therefore, $\sigma_{tc}$ is proportional to 
$s/\Lambda^4$ (see (\ref{const})) and grows with the c.m. energy for 
a fixed $\Lambda$. 
Clearly, for this approximation to be valid, $\Lambda$ must be 
larger than $\sqrt s$. 

A few more useful 
observations can be made already by looking at the polarized 
cross sections in (\ref{sigll})--(\ref{sigrr}) above:
\begin{itemize} 
\item There are no interference effects between the 
different four-Fermi couplings $V_{ij},~S_{RR}$ 
and $T_{RR}$; the total cross section depends 
only on the square value of these couplings and 
is, therefore, maximal when all these couplings are non-zero. 
\item The vector couplings appear in the total cross section
only in the combination $ \sum | V_{ij}|^2 $. 
\item Initial and/or final 
polarization of the incoming electrons 
and/or top quarks can distinguish between different sets of 
couplings, e.g., if the 
incoming electron beam if left polarized then only $V_{LL}$ and 
$V_{LR}$ can contribute to $t \bar c$ production. 
\end{itemize} 

Before continuing we note
that the new $Zt c$ couplings $a_L^Z$ and $a_R^Z$ in 
(\ref{ztc}) also contribute to $e^+e^- \to t \bar c$ by interfering 
with the four-Fermi vector couplings $V_{ij}$.
These effects can be included by redefining
\begin{eqnarray}
V_{ij} \to V_{ij} + 4 c_i^Z a_j^Z \frac{m_W m_Z}{s-m_Z^2}~,
\end{eqnarray}
where $i,j=L,R$, and 
$c_L^Z=-1/2+s_W^2$, $c_R^Z=s_W^2$ are the couplings of a Z-boson to 
a left or a right handed electron, respectively.
The effects of such new $Ztc$ vector couplings on 
$e^+e^- \to t \bar c$ 
were also recently investigated by Han and Hewett \cite{joan}, 
who have made a detailed analysis 
of the sensitivity of 200 - 1000 GeV $e^+e^-$ colliders 
to such new couplings. 
Here, for the process $e^+e^- \to t \bar c$, 
we instead focus mainly on the effects of the 
four-Fermi couplings which, as will be shown below, give the 
dominant contribution to $\sigma_{tc}$. 

\begin{figure}[htb]
\psfull
 \begin{center}
 \leavevmode
 \epsfig{file=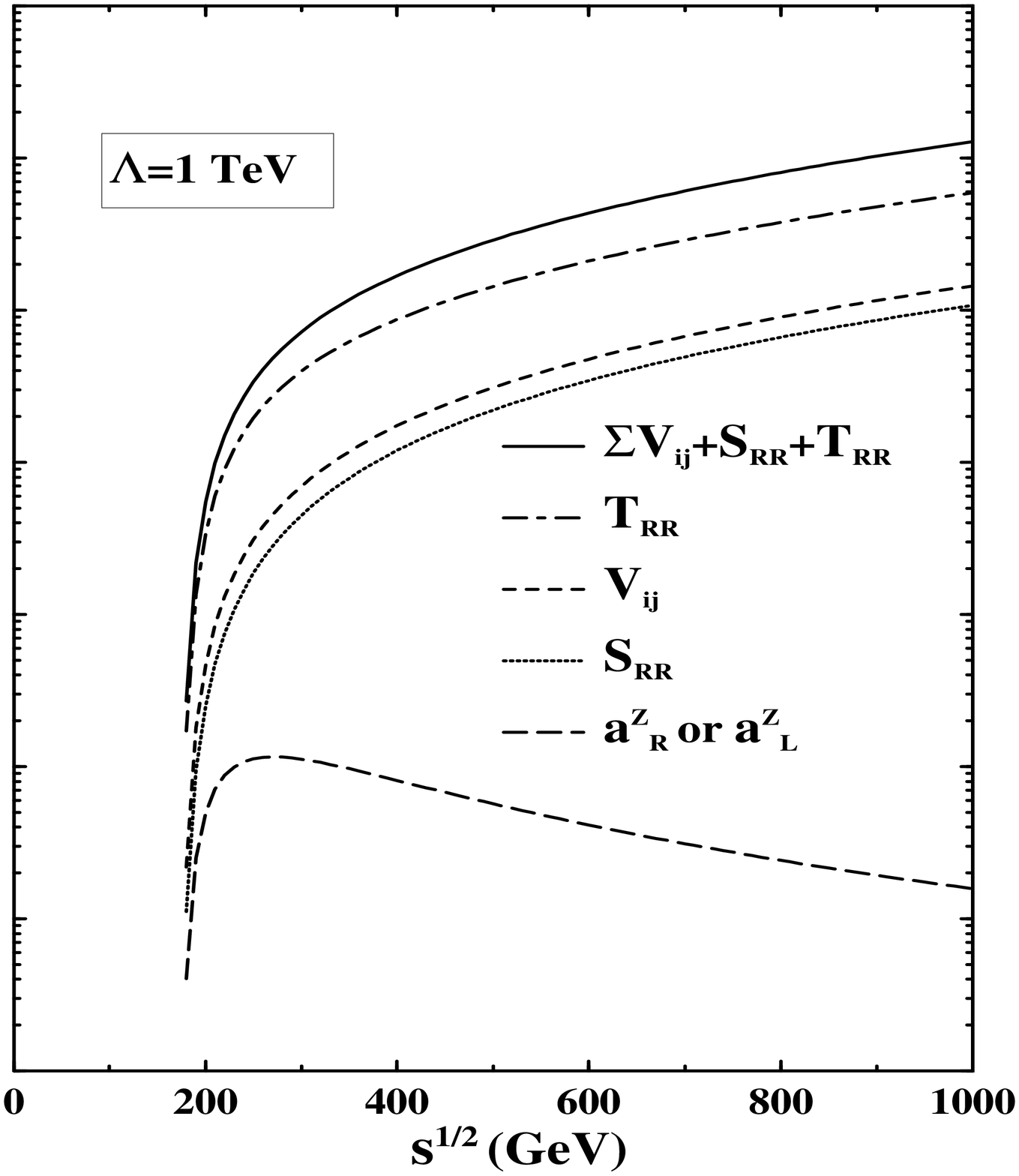,height=8cm,width=8cm,bbllx=0cm,bblly=2cm,bburx=20cm,bbury=25cm,angle=0}
 \end{center}
\caption{\emph{The cross section 
$\sigma_{tc}=\sigma(e^+e^- \to t \bar c + \bar t c)$ (in $fb$) is plotted as a 
function of the c.m. energy ($\sqrt s$) of the $e^+e^-$ collider.
The following cases are shown: all four-Fermi couplings are non-zero
and equal 1, i.e., $ V_{LL}= V_{LR}=
 V_{RL}= V_{RR}= S_{RR}= T_{RR}=1$ (solid line), 
only $ T_{RR}=1$ (dot-dashed line), only one of the vector couplings 
$ V_{ij}$ equals 1 (dashed line), only $ S_{RR}=1$ (dotted line)
and either $ a_L^Z=1$ or $ a_R^Z=1$ with the four-Fermi couplings 
set to zero (long-dashed line). $\Lambda=1$ TeV is used for all cases.}}
\label{eetcfig1}
\end{figure}

In Fig.~\ref{eetcfig1} we plot the total cross section $\sigma_{tc}$ (in $fb$)
as a function of the c.m. energy of the $e^+e^-$ collider, taking 
$\Lambda=1$ TeV, and for different types of four-Fermi couplings;
as expected, the four-Fermi effective couplings give contributions to
$\sigma_{tc}$ which grow with the c.m. energy. Due to this effect, the
cross section can be rather 
large, ranging from about 30 $fb$ to 300 $fb$ 
and yielding tens to hundreds $ t \bar c$ events
(depending on the type of four-Fermi coupling) 
already at LEP2 energies.
 At a 1 TeV NLC we find that 
$\sigma_{tc} \sim 10^4 - 10^5$ $fb$ if $\Lambda=1$ TeV. Recall 
that $\sigma_{tc}$ scales as $1/\Lambda^4$, therefore, even with 
$\Lambda \sim 10 - 20$ TeV $\sigma_{tc}$ is of ${\cal O}(fb)$
at a 1 TeV NLC. 
 
For completeness we also plot $\sigma_{tc}$ for non-zero 
$Ztc$ couplings $a_L^Z=1$ or $a_R^Z=1$ (dashed line). 
Clearly, the effects of such couplings are much smaller than 
those generated by the four-Fermi interactions. Even 
at LEP2 the contribution is about 
one to two orders of magnitudes smaller than the typical 
contribution from the four-Fermi interactions. 

Notice also that, contrary to the four-Fermi case, the $Ztc$
contributions to $\sigma_{tc}$ drop as $\sim 1/s$ 
due to the explicit $s$-channel $Z$-boson propagator.
Because of this, at a NLC with c.m. energies of $\sqrt s \gsim 1.5$ TeV, 
$t$-channel vector-boson fusion processes $W^+W^- \to t \bar c$ 
(see Fig.~\ref{fig3}(a)) and $ZZ \to t \bar c$ 
(see Fig~\ref{fig4}(a) and (b)) become important and 
may be better probes of such $Ztc$ couplings.
We have calculated the total cross sections 
$\sigma_{WW} = \sigma(e^+e^- \to W^+W^- \nu_e \bar\nu_e \to 
t \bar c \nu_e \bar\nu_e)$ and $\sigma_{ZZ} = 
\sigma(e^+e^- \to ZZ e^+ e^- \to 
t \bar c e^+ e^-)$ 
using the effective vector boson approximation (EVBA) \cite{evba}.
In this approximation, 
as in the equivalent photon approximation in QED, the colliding 
$W$'s or $Z$'s are treated as on shell particles and, 
thus, the salient features of the $2 \to 4$ 
reactions $e^+e^- \to t \bar c \nu_e \bar\nu_e,~t \bar c e^+e^-$ 
are generated by the simpler $2 \to 2$ 
sub-processes $W^+W^-, ~ ZZ \to t \bar c$. The full $2 \to 4$ 
cross sections $\sigma_{V_1V_2}$ ($V_1,V_2=W^+,W^-$ or $V_1,V_2=Z,Z$) are 
estimated by folding in the 
distribution functions $f_{V_1}^{\lambda_1},f_{V_2}^{\lambda_2}$ 
of the two colliding $V_1,V_2$ with helicities $\lambda_1,\lambda_2$
\cite{evba}, explicitly,
\begin{eqnarray}
\sigma_{V_1V_2} = \sum_{\lambda_1,\lambda_2} \int dx_1 dx_2 
f_{V_1}^{\lambda_1}(x_1) f_{V_2}^{\lambda_2} (x_2) 
\hat \sigma(V_1^{\lambda_1} V_2^{\lambda_2} \to t \bar c) 
\label{fullcross}~.
\end{eqnarray}
We find that $\sigma_{ZZ} \lsim 10^{-3}$ $fb$ at 
$\sqrt s = \Lambda = 1.5$ TeV, 
for $ a_L^Z=1$ or $ a_R^Z=1$, and is therefore too small to be observed. 
However, $\sigma_{WW}$ is typically about two orders of magnitude larger, 
partly because in this approximation, 
the $W$-boson luminosity is larger 
than the luminosity for the $Z$-bosons 
due to different couplings to electrons (see e.g., \cite{collider}).
In particular, we find $\sigma_{WW} \sim 0.15 ~(0.09)$ $fb$ at
$\sqrt s = \Lambda = 1.5~(2)$ TeV, 
for $a_L^Z=1,~a_R^Z=0$ or $a_L^Z=0,~a_R^Z=1$. Comparing with 
$\sigma(e^+e^- \to Z \to t \bar c + \bar t c) \sim 0.14 ~(0.03)$ $fb$ 
for the same values
of $\sqrt s ,~ \Lambda $ and $ a_{L,R}^Z $, 
we see that the $WW$-fusion process is a slightly more sensitive
probe of such new $Ztc$ couplings at these
high c.m. energies.\footnote{We recall that, at these high c.m. 
energies ($\sqrt s = 1.5 -2$ TeV), the 
projected integrated luminosity is expected to be several 
hundreds inverse $fb$, see Ref.~\cite{nlcreviews}. Thus, 
a cross section of the order of 0.1 $fb$ may yield an observable effect, 
especially for the rather unique $t \bar c$ final state which 
has a negligible background as we discuss below.} 
 
Let us now return to the four-Fermi case;
we wish to explore the limits that can be obtained on the scale $\Lambda$ 
of such four-Fermi operators in the case that no $e^+e^- \to t \bar c$ 
events are observed. 
To do so we first consider the possible observable 
final states for this reaction:

\begin{enumerate}
\item If the top decays hadronically via $t \to bW^+ \to b j_1j_2$, 
where $j_1,j_2$ are light jets coming from $W^+ \to u \bar d~{\rm or}~
c \bar s$, then 
we have $e^+e^- \to t \bar c \to b \bar c j_1j_2$ 
(and $e^+e^- \to \bar t c \to \bar b c \bar j_1 \bar j_2$ 
for the charge conjugate channel). These final states occur with 
a branching ratio of 2/3. 

\item If the top decays semi-leptonically via $t \to bW^+ \to b \ell^+ 
\nu_\ell$, 
where $\ell=e,\mu$ or $\tau$, then 
we have $e^+e^- \to b \bar c \ell^+ \nu_\ell$ 
(and $e^+e^- \to \bar b c \ell^- \bar\nu_\ell$ 
for the charge conjugate channel). These final states occur with 
a branching ratio of 1/3. 

\end{enumerate}

An immediate useful observation is that each of the two top decay 
scenarios above contains a single $b$-jet in the final state, which can
be used as a signal for non-SM physics \cite{bparity}. Indeed, 
SM reactions in lepton colliders produce almost exclusively final states with
an {\em even} number of $b$-jets. Defining a quantum
number we called $b_P = (-1)^n$, where $n$ is the number of
$b$-jets in the final state, the SM is almost
exclusively $b_P$-even. The SM irreducible 
background to $b_P$-odd processes generated by new physics
is severely suppressed by off-diagonal CKM elements 
and can be neglected. The only remaining (reducible) 
background to processes which yield an odd number of $b$-jets 
in the final state arises from miss-identifying an odd number 
of $b$-jets in a $b_P$-even event \cite{bparity}.

For the process $e^+e^- \to t \bar c$, the SM irreducible
background is generated, for example, by $e^+e^- \to W^+W^-$ followed by
$W^+ \to j_1j_2$ and $W^- \to b \bar c$ 
for hadronic top decays (case 1 above) or by 
$W^+ \to \ell^+ \nu_\ell$ and $W^- \to b \bar c$ 
for semi-leptonic top decays (case 2 above), see also \cite{joan}.
These backgrounds are clearly CKM suppressed, 
being $\propto |V_{cb}|^2$, and can therefore be neglected. 

In order to further eliminate the reducible background to
$b_P$-odd events produced by a $b$-tagging efficiency below 1,
one can employ a few more specific experimental handles allowed by the very 
distinct characteristics of 
a $t \bar c$ signature: {\it(i)} The possibility of efficiently 
reconstructing the $t$ from the decay $ t \to bW \to b j_1 j_2 $ at the 
NLC \cite{frey}; the top quark can also be reconstructed 
in the case of semi-leptonic top decays 
since there is only one missing neutrino in such a $t \bar c$ event.
{\it(ii)} Since this is a $2 \to 2$ process, the two-body kinematics 
fixes the charm-jet energy to be $E_c \simeq \sqrt s (1-m_t^2/s)/2$. 
The charm-jet gives then a unique signal since it recoils against 
the massive top quark and should stand out as a very energetic 
light jet at high c.m. energies. The event will then
look like a single top quark event.
{\it(iii)} The energy of the $b$-jet produced in top decay is 
also known due to two-body kinematics \cite{joan}. 

Let us therefore define our background-free observable cross section, 
which we denote by $\bar\sigma_{tc}$, as 
the effective cross section including $b$-tagging efficiency 
($\epsilon_b$) and top quark reconstruction efficiency 
($\epsilon_t$)
\begin{eqnarray}
\bar\sigma_{tc} &=& \epsilon_b \epsilon_t \sigma_{tc} \label{sigtcbar}~.
\end{eqnarray}

We define the largest $\Lambda$ to which a collider is
sensitive as the one for which 
10 fully reconstructed $t \bar c + \bar t c$ events are
generating per year, after eliminating any potential background,
i.e., the value of $\Lambda$ for which $\bar\sigma_{tc} \times L = 10$, 
where $L$ is the yearly integrated luminosity of the given collider.

In Table 1 we list the 
limits that can be placed on the scale $\Lambda$ of the new 
effective four-Fermi and $Ztc$ operators, 
based on this 10 event criterion, using 
the background-free cross section as defined in (\ref{sigtcbar}); we take 
$\epsilon_b=60\%$ and $\epsilon_t=80\%$ and we impose a 
$10^\circ$ angular cut on the c.m. scattering angle.\footnote{We note 
that our limits on the scale of the $Ztc$ operator are more stringent 
than those obtained in \cite{joan}. This difference arises from our 
assumption that once the top quark is reconstructed (with an efficiency of 
$\epsilon_b \times \epsilon_t$) and the charm jet is identified (as 
described above), there is no additional background to be 
considered for the $t c$ final state; the results of \cite{joan}
obtained using a more careful background estimate
correspond to a reduced reconstruction efficiency of $
\epsilon_b \epsilon_t = 42\% $ instead of $48\%$ which we used.} 
The limits are calculated, assuming that 
only one coupling is non-vanishing at a time, i.e., 
with either $ V_{ij}=1$, for $i,j=LL$ or $LR$ or $RR$ or $RL$, or 
$ S_{RR}=1$ or $ T_{RR}=1$ or $ a_L^Z=1$. 
We give the limits that may already be obtainable from 
the recent 189 GeV run of LEP2 which accumulated $\sim 150$ inverse $pb$
in each of the four LEP2 detectors \cite{shen}.   
We also consider 
three future collider scenarios: LEP2 with 
a c.m. energy of $\sqrt s=200 $ GeV and an integrated luminosity 
of $L=2.5$ $fb^{-1}$,
a NLC with $\sqrt s=500 $ GeV and $L=50$ $fb^{-1}$ and 
a NLC with $\sqrt s=1000 $ GeV and $L=200$ $fb^{-1}$.
As expected, the strongest limits are obtained using
the four-Fermi couplings. 
In particular, assuming that no $t \bar c$ event was seen during 
the recent LEP2 run, this rules out new flavor physics (that can generate
such four-Fermi operators) up to energy scales of  
$\Lambda \gsim 0.7-1.4$ TeV.  
For the future $e^+e^-$ machines, 
the limits on the scale of the four-Fermi operators 
are typically $\Lambda \gsim 7-12 \times \sqrt s$ for LEP2 energies and 
$\Lambda \gsim 17-27 \times \sqrt s$ for a 500 or 1000 GeV NLC.
The best limits are obtained on the tensor four-Fermi coupling 
$T_{RR}$ due to numerical factors in the cross section. \\

$$
\begin{tabular}{||c|c|c|c|c|c||} 
\hline
\multicolumn{6}{||c||}{Limits from $\bar\sigma_{tc}=\epsilon_b \epsilon_t 
\sigma(e^+e^- \to t \bar c)$}\\ \hline 
{}&
{}&
{$ a_i^Z=1$}&
{$ V_{ij}=1$}&
{$ S_{RR}=1$}&
{$ T_{RR}=1$}\\ 
$\sqrt{s}$ & 
$L$&
{$i=L$ or $R$}&
{$ij=LL,LR,RR$ or $RL$}&
{}&
{}\\ 
\hline
189~GeV&
0.6~$fb^{-1}$&
0.5~TeV&
0.8~TeV&
0.7~TeV&
1.4~TeV
\\ \hline
200~GeV&
2.5~$fb^{-1}$&
0.9~TeV&
1.5~TeV&
1.3~TeV&
2.5~TeV
\\ \hline
500~GeV&
50~$fb^{-1}$&
1.9~TeV&
9.3~TeV&
8.5~TeV&
13.6~TeV
\\ \hline
1000~GeV&
200~$fb^{-1}$&
2.0~TeV&
19.3~TeV&
17.9~TeV&
27.5~TeV
\\ 
\hline
\end{tabular}
$$

\bigskip
\bigskip

{\bf Table 1:} {\emph {The limits on the scale of the new physics $\Lambda$ 
using the reaction $e^+e^- \to t \bar c + \bar t c$.
The limits are given for one non-vanishing coupling 
at a time and setting this coupling to $1$.
In each case four accelerator scenarios are considered;
$\sqrt{s}=189$, $200$, $500$ and $1000$ GeV with luminosities
$L=0.6$, $2.5$, $50$ and $200~fb^{-1}$, respectively. 
The signals considered are based on the total cross section as defined 
in (\ref{sigtcbar}), 
assuming a $b$-tagging efficiency of $60\%$ and a top reconstruction 
efficiency of $80\%$ (see text). 
Also, the limits are based on the criterion of $10$ events 
for the given luminosity.}}

\bigskip
\bigskip

The above results were obtained assuming
that all couplings were equal to 1, for other values the 
the limits in Table 1 are in fact on $\Lambda/\sqrt {f}$, where 
$f=V,~S,~T$ or $a^Z$. To illustrate this possibility
consider, for example, the tensor
four-Fermi coupling which can (of course) be generated by the
exchange of a heavy neutral tensor excitation, 
of mass $\Lambda $. But this effective vertex is also
generated through Fierz-transforming the operator 
$ {\cal O}_{\ell q^\prime}$ in (\ref{4fermi7}), 
which can be produced
by the exchange of a heavy scalar leptoquark in the underlying 
high energy theory. In the latter case the coefficient $T_{RR}$ has an
additional factor of $1/8$, so that the mass of the leptoquark
corresponds to $\sqrt{8} \Lambda $. These two possibilities cannot be 
easily differentiated using an effective theory and provide an example 
of the limitations of this parameterization.

It is also instructive to note 
that, in case a $t \bar c$ signal is observed, 
there are enough independent observables in the reaction
$e^+e^- \to t \bar c$ to allow the extraction of all
6 independent four-Fermi couplings discussed above. These observables
are, for example, the cross sections for polarized electrons and for definite top
polarization (viable in the semi-leptonic \cite{peskin} and 
in the hadronic top decays if the down quark jet 
can be distinguished from the up quark jet in $W \to du$ \cite{topspinhad}), 
and the following forward-backward (FB) asymmetries 
for polarized incoming electrons (i.e., for the 
reactions $e^-_L e^+ \to t \bar c$ and $e^-_R e^+ \to t \bar c$)
\begin{eqnarray}
A_{{FB}_L} &=& \frac{\int_0^{\pi/2} \left\{ d\sigma_{e_L t_L}(\theta) + 
d\sigma_{e_L t_R}(\theta) - d\sigma_{e_L t_L}(\pi - \theta) - 
d\sigma_{e_L t_R}(\pi - \theta) \right\}}{\sigma_{e_L t_L}+ 
\sigma_{e_L t_R}} \nonumber \\
&=& \frac{3 (1+\beta_t)}{2 (3+\beta_t)} 
\frac{V_{LR}^2 - V_{LL}^2}{ V_{LR}^2 + V_{LL}^2} 
\label{afbl}~,\\
~\nonumber \\
A_{{FB}_R} &=& \frac{\int_0^{\pi/2} \left\{ d\sigma_{e_R t_L}(\theta) + 
d\sigma_{e_R t_R}(\theta) - d\sigma_{e_R t_L}(\pi - \theta) - 
d\sigma_{e_R t_R}(\pi - \theta) \right\}}{\sigma_{e_R t_L}+ 
\sigma_{e_R t_R}} \nonumber\\
&=& \frac{3 (1+\beta_t) \left[ V_{RL}^2 - V_{RR}^2 +
4 S_{RR} T_{RR} \right] }{
2 (3+\beta_t) \left[ V_{RL}^2 + V_{RR}^2 +
3 (1+\beta_t) S_{RR}^2 + 16(3-\beta_t) T_{RR}^2 \right] } 
\label{afbr}~.
\end{eqnarray}
Clearly, the FB asymmetries involve ratios of cross sections 
and, therefore, are not suppressed by inverse powers of $\Lambda$. 
A detailed discussion of how to extract the six four-Fermi couplings 
from such observables lies
outside the scope of this paper, we limit ourselves to the summary
of the sensitivity of each observables as presented in Table 2.\\

$$
\begin{tabular}{||c||c|c|c|c|c|c||} 
\hline
\multicolumn{7}{||c||}{Observables Vs. Couplings}\\ \hline \hline
{}&
{$\sigma_{e_L t_L}$}&
{$\sigma_{e_L t_R}$}&
{$\sigma_{e_R t_L}$}&
{$\sigma_{e_R t_R}$}&
{$A_{{FB}_L}$}&
{$A_{{FB}_R}$}\\
\hline \hline
$ V_{LL}$ & 
$\surd$&
$\surd$&
{}&
{}&
$\surd$&
{}\\ 
\hline
$ V_{LR}$ &
$\surd$&
$\surd$&
&
&
$\surd$&
{} \\ 
\hline
$ V_{RR}$ &
&
&
$\surd$&
$\surd$&
{}&
$\surd$\\ 
\hline
$ V_{RL}$ &
&
&
$\surd$&
$\surd$&
{}&
$\surd$\\ 
\hline
$ S_{RR}$ &
&
&
$\surd$&
&
{}&
$\surd$\\ 
\hline
$ T_{RR}$ &
&
&
$\surd$&
$\surd$&
{}&
$\surd$\\ 
\hline
\end{tabular}
$$

\bigskip
\bigskip

{\bf Table 2:} {\emph {The sensitivity of the different observables 
discussed in the text to the various new four-Fermi effective couplings.
The observables considered are the
polarized cross sections 
$\sigma_{e_i t_j}$, $i,j=LL,~LR,~RR,~RL$ in (\ref{sigll})--(\ref{sigrr})
and the FB asymmetries $A_{{FB}_L}$ and $A_{{FB}_R}$ for left and 
right-handed incoming electron beam, respectively, as defined in 
(\ref{afbl}) and (\ref{afbr}). 
A check-mark shows that the given observable is sensitive to the 
given coupling.}}

\bigskip
\bigskip 

We conclude this section with a few remarks.

\begin{itemize}

\item Some of the four-Fermi effective operators 
in (\ref{4fermi1})--(\ref{4fermi7}), that 
generate the new $tcee$ coupling
also induce interactions involving the down quarks 
of the second and third generations. For example, the operators 
${\cal O}_{\ell q}^{(1)}$ and ${\cal O}_{\ell q}^{(3)}$, being 
constructed out of the left-handed quark doublets, will generate 
a $t \bar c e^+e^- $ interaction (with coupling $ \alpha_{\ell q}^{(1)} -
 \alpha_{\ell q}^{(3)} $, see (\ref{vst})) as well as a 
$b \bar s e^+e^-$ one with coupling $ \alpha_{\ell q}^{(1)} +
 \alpha_{\ell q}^{(3)} $.
This fact, a consequence of gauge invariance,
can be used to derive constraints on the scale $ \Lambda $. For example, 
using the measured $B^+$ semi-leptonic branching ratio, 
such four-Fermi operators contributions to 
$B^+ \to K^+ e^+ e^- $ will be below the existing bound 
Br($B^+ \to K^+ e^+ e^- $)$<10^{-5}$ \cite{pdg}, 
provided that $ \Lambda/\sqrt{ |\alpha_{\ell q}^{(1)} +
 \alpha_{\ell q}^{(3)}|}  \gsim 2 $ TeV. 
Due to the different 
combination of couplings
appearing in this expression this bound is complementary to the ones
obtained above.

\item We wish to emphasize the importance of adding such possible 
four-Fermi 
interactions to a model independent analysis of $e^+e^- \to t \bar c$.
We argued previously that the only models that can produce observable
flavor violations are those which generate flavor-changing operators at
tree-level. In this case the effective $Ztc$ vertex is generated
by the exchange of a heavy
gauge boson $V$ which mixes with the $Z$ and which has a $Vtc$ vertex
(this vertex is also produced by heavy fermion exchanges). 
Similarly, some of the four-Fermi operators are generated by the exchange
of a heavy vector $ V' $ coupling to $ t \bar c $ and to $ e^+ e^- $. In
general we have $ V' \not = V $ so that an analysis that covers all the
possibilities allowed by an effective Lagrangian parameterization should
include both types of vertices. If, on the other hand $ V = V' $, then
the bounds obtained form the four-Fermi contact interactions are far
superior to the ones derived from $Z$-mediated reactions. In this case
$V$ would also generate a $eeee$ contact interaction for which existing
limits \cite{pdg} give $ \Lambda \gsim 1.5 $ TeV, i.e., better 
than the limits given in Table 1 for a 189 GeV LEP2. We note, however, 
that at a 500 GeV NLC the limits 
that can be obtained on the $eeee$ contact terms 
by studying the reaction $e^+e^- \to e^+e^-$ are about $\Lambda > 5$ TeV
\cite{jose2}, while from Table 1 we see that $\Lambda > 8.5 - 13.6$ TeV
is attainable at this energy by studying the process $e^+e^- \to t \bar c$.

\item We would like to stress again that the limits obtained
in Table 1 presuppose the 
heavy physics does generate the four-Fermi operators at an accessible scale.
Other types of new physics can be responsible for 
generating the $Ztc$ vertex, raising the possibility that the latter 
occurs even when the former is negligible. In that sense, the above results 
are complementary to those obtained e.g., in \cite{prl82p1628}.

\end{itemize}

\section{Effective flavor-changing scalar interactions and 
\boldmath{$t \bar c$} production at a NLC}

In this section we consider neutral Higgs exchanges in the NLC 
which lead to $t \bar c$ production via 
a new $ht c$ interaction as defined in section 2
(see (\ref{htc}) and (\ref{arh})).
We Neglect $2 \to 3$ (i.e., three-body final state) processes since 
these are suppressed by 
phase space compared to $2 \to 2$ processes. 

There are only two such reactions that can probe an effective 
$ht c$ vertex in $e^+e^-$ colliders. 
The first is the Bjorken process 
$e^+e^- \to Zh$ when a real Higgs is produced and then decay 
to a $t \bar c$ pair (see Fig.~\ref{fig2}), and the second is the 
$t$-channel $W^+W^-$-fusion to a neutral Higgs in Fig.~\ref{fig3}(b), 
leading to
$e^+e^- \to W^+ W^- \nu_e \bar\nu_e \to t \bar c \nu_e 
\bar\nu_e$. We note that the corresponding $t$-channel 
$ZZ$-fusion process $e^+e^- \to ZZ e^+e^- \to t \bar c e^+ e^-$,
also depicted in Fig.~\ref{fig3}(b), 
is about an order of magnitude smaller 
than the $WW$-fusion process, basically, due to the different 
couplings of a $Z$-boson to electrons (see also the discussion in the previous 
section).

We focus on Higgs masses in the range $m_t \lsim m_h \lsim 500$ GeV. 
Since at this mass range the neutral Higgs width is still quite small 
compared to its mass, e.g., for $m_h=250(500)$ GeV 
the width is about $1.5\% (13\%)$ of its mass, 
and since we only consider real Higgs production, 
we may estimate the cross section for $e^+e^- \to Zt \bar c$ by
\begin{eqnarray}
\sigma_{Ztc} = \sigma( e^+e^- \to Zt \bar c) 
\approx \sigma(e^+e^- \to Zh) \times {\rm Br}(h \to t \bar c)~,
\end{eqnarray} 
where \cite{collider}
\begin{eqnarray}
\sigma(e^+e^- \to Zh) = \frac{\pi \alpha^2}{192 c_W^4 s_W^4}
\left[ 1+(1-4 s_W^2)^2\right]
\frac{8 \kappa}{\sqrt s}
{ \left(\kappa^2+3m_Z^2 \right) \over \left( s - m_Z \right)^2 }~,
\end{eqnarray}
and
\begin{eqnarray}
\kappa = \sqrt{\frac{(s+m_Z^2-m_h^2)^2 - 4sm_Z^2}{4 s}}~.
\end{eqnarray} 
Using now
\begin{eqnarray}
{\rm Br}(h \to t \bar c) = \frac{\Gamma(h \to t \bar c)}{\Gamma_h} ~,
\end{eqnarray}
where $\Gamma_h=\Gamma(h \to b \bar b) + \Gamma(h \to ZZ) + 
\Gamma(h \to W^+W^-) + \Gamma(h \to t \bar t)$ is the total SM
Higgs width\footnote{$h\to t \bar t$ and $h \to ZZ$ are 
included when kinematically allowed.}
, see e.g., \cite{higgshunters},
and $\Gamma(h \to t \bar c)$ is calculated
in terms of the new $ht c$ couplings 
$a_L^h$ and $a_R^h$ defined in (\ref{htc}); we find 
\begin{eqnarray}
\Gamma(h \to t \bar c)=\frac{v^4}{\Lambda^4} \frac{3 \alpha}{4 s_W^2} 
\left[ (a_L^h)^2 + (a_R^h)^2 \right] m_h
\left(1 - \frac{m_t^2}{m_h^2} \right)^2 ~.
\end{eqnarray}

\begin{figure}[htb]
\psfull
 \begin{center}
 \leavevmode
 \epsfig{file=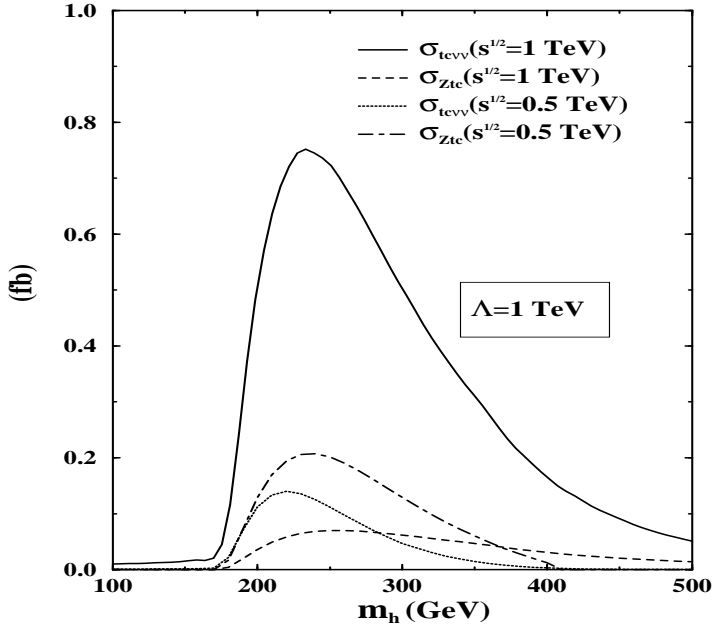,height=8cm,width=8cm,bbllx=0cm,bblly=2cm,bburx=20cm,bbury=25cm,angle=0}
 \end{center}
\caption{\emph{The cross sections 
$\sigma_{Ztc}=\sigma(e^+e^- \to Zt \bar c + Z\bar t c)$ 
and $\sigma_{tc \nu \nu}=\sigma(e^+e^- \to t \bar c \nu_e \bar\nu_e + 
\bar t c \nu_e \bar\nu_e)$
(in $fb$) are plotted as a 
function of the SM Higgs mass $m_h$, for an $e^+e^-$ collider 
with a 
c.m. energy of $\sqrt s=500$ GeV (dotted and dot-dashed lines) and 
of $\sqrt s=1000$ GeV (solid and dashed lines).
$\Lambda=1$ TeV is used for all cases.}}
\label{htcfig1}
\end{figure}

\noindent For this type of effective vertices we also calculate
the $t$-channel fusion cross section $\sigma_{tc\nu \nu} 
= \sigma (e^+e^- \to t \bar c \nu_e \bar\nu_e)$
using the EVBA \cite{evba} (see also the previous section).
The amplitude for the hard  $2 \to 2$ sub-process 
$W^+_{\lambda^+} W^-_{\lambda^-} \to h \to t \bar c$ 
with c.m. energy $ \sqrt{\hat s} $ is given by 
\begin{eqnarray}
{\cal M}_{\lambda^+,\lambda^-}&=& \frac{v^2}{\Lambda^2} 
\frac{\pi \alpha}{s_W^2} \hat s 
\sqrt{1-\beta_W^2} \sqrt{\frac{2 \beta_t}{1+\beta_t}} \Pi_h 
\delta_{\lambda_t,\lambda_c} T_{\lambda^+,\lambda^-} \nonumber\\
&&~~~~~~~~~~\times \left[ a_L^h(1+\lambda_t) - a_R^h(1-\lambda_t) \right]~,
\end{eqnarray}
where $\lambda^+,\lambda^-=0,~\pm 1$ are the helicities of the
$W^+,W^-$, respectively, and $ \lambda_{t,c} = \pm 1/2 $ denote the quark
helicities. Also, 
$\Pi_h= (\hat s - m_h^2 + i m_h \Gamma_h)^{-1}$ is the 
Higgs propagator, $\beta_t=(\hat s - m_t^2)/(\hat s + m_t^2)$, 
$\beta_W = \sqrt{1- 4m_W^2/\hat s}$ and
\begin{eqnarray}
T_{0,0} = \frac{1+\beta_W^2}{1-\beta_W^2} ~~,~~
T_{\pm,\pm}=1~~,~~T_{\pm,\mp}=T_{\pm,0}=T_{0,\pm}=0~.
\end{eqnarray} 
The polarized (with 
respect to the $W^+$ and $W^-$) hard cross section 
$\hat\sigma(W^+_{\lambda^+} W^-_{\lambda^-} \to h \to t \bar c)$ can then be 
readily calculated. From this expression the
cross section $\sigma_{tc\nu \nu}$ is again estimated by folding in the 
distribution functions $f_{W^+}^{\lambda^+},f_{W^-}^{\lambda^-}$ of the two 
colliding $W^+,W^-$ in a given helicity state $\lambda^+,\lambda^-$ 
as in (\ref{fullcross})
\begin{eqnarray}
\sigma_{t c \nu \nu} = \sum_{\lambda^+ \lambda^- } \int dx_+ \, dx_- \,
f_{W^+}^{\lambda^+}(x_+) f_{W^-}^{\lambda^-} (x_-) 
\hat\sigma(W^+_{\lambda^+} W^-_{\lambda^-} \to h \to t \bar c)~.
\end{eqnarray}
The bulk contribution to the 
full $2 \to 4$ process arises when the Higgs resonates, i.e., 
when $\sqrt{\hat s} \sim m_h$
($\sqrt{\hat s}$ denotes the c.m. energy of the hard $2\to2$ process). 
Because of this, ${\rm Br}(h \to t \bar c)$ 
also controls the dependence 
of this $WW$-fusion reaction on the Higgs mass. 
This behavior is illustrated in Fig.~\ref{htcfig1} in which we plot 
$\sigma_{Ztc}$ and 
$\sigma_{tc\nu \nu}$ 
as a function of $m_h$ for c.m. energies of 500 and 1000 GeV. 
In this figure we take $ a_L^h= a_R^h=1$ and 
$\Lambda=1$ TeV. 
We see that these cross sections reach their maximum for
$m_h \sim 230$ GeV, close to the value at which ${\rm Br}(h \to t \bar c)$
is largest.

\begin{figure}[htb]
\psfull
 \begin{center}
 \leavevmode
 \epsfig{file=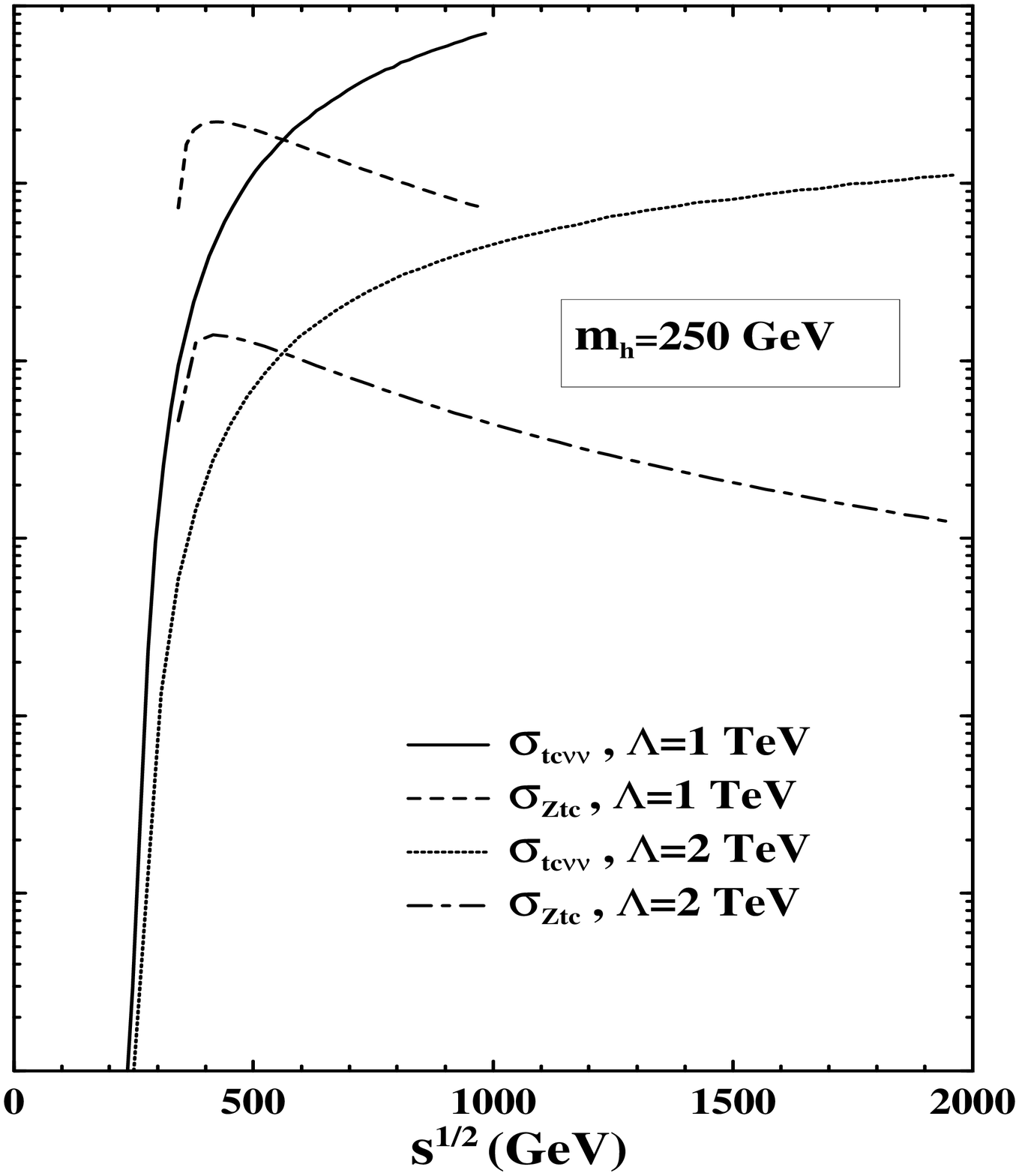,height=8cm,width=8cm,bbllx=0cm,bblly=2cm,bburx=20cm,bbury=25cm,angle=0}
 \end{center}
\caption{\emph{The cross sections 
$\sigma_{Ztc}=\sigma(e^+e^- \to Zt \bar c + Z\bar t c)$ 
and $\sigma_{tc \nu \nu}=\sigma(e^+e^- \to t \bar c \nu_e \bar\nu_e + 
\bar t c \nu_e \bar\nu_e)$
(in $fb$) are plotted as a
function of the c.m. energy of the $e^+e^-$ collider, for $m_h=250$ GeV
and for: $\Lambda=1$ TeV (solid and dashed lines) and 
$\Lambda=2$ TeV (dotted and dot-dashed lines). See also text.
}} 
\label{htcfig2}
\end{figure}

We now discuss these signals and their observability 
in future high energy $e^+e^-$ colliders. 
From Fig.~\ref{htcfig1} we see that, at c.m. energy of 
500 GeV, the Bjorken process dominates, 
giving $\sigma_{Ztc} \sim 0.2$ $fb$ for $m_h \sim 250$ GeV, 
$ a_L^h= a_R^h=1$ and $\Lambda=1$ TeV. 
In Fig.~\ref{htcfig2} we plot $\sigma_{Ztc}$ and 
$\sigma_{tc\nu \nu}$ as a function of 
the $e^+e^-$ c.m. energy $\sqrt s$, for $m_h=250$ GeV, $ a_L^h= a_R^h=1$ and 
$\Lambda=1$ or 2 TeV.\footnote{Notice that some of the lines end rather 
abruptly whenever $\sqrt s = \Lambda$, since for $\Lambda < \sqrt s$ the 
effective Lagrangian description is not valid by definition.}
Since $\sigma_{tc\nu \nu}$ is a $t$-channel fusion process, it 
grows logarithmically as $\sim {\rm log}^2(s/m_W^2)$ and, therefore, 
dominates at higher energies over the $s$-channel 
Bjorken process which drops as $ \sim 1/s$. For example, 
at $\sqrt s =1$ TeV and $m_h \sim 250$ GeV we find
$\sigma_{tc \nu \nu }/\sigma_{Ztc} \sim 10$. 

In order to identify the background to these reactions, we follow the
same approach described in the previous section. We consider
the possible observable final states in $e^+e^- \to Z t \bar c$ 
(assuming that the $Z$ is identified with $100\%$ efficiency) and 
$e^+e^- \to t \bar c \nu_e \bar\nu_e$, which are determined 
by the top decays. For hadronic top decays we have 
$e^+ e^- \to Z t \bar c \to Z \bar c b j_1 j_2$ and 
$e^+e^- \to t \bar c \nu_e \bar\nu_e \to \bar c b j_1 j_2 \nu_e \bar\nu_e$, 
where $j_1$ and $j_2$ are light jets from 
$W^+ \to u \bar d$ or $W^+ \to c \bar s$. For the semi-leptonic 
top decays we have $e^+ e^- \to Z t \bar c \to Z \bar c b \ell^+ \nu_\ell$ 
and $e^+e^- \to t \bar c \nu_e \bar\nu_e \to \bar c b \ell^+ \nu_\ell 
\nu_e \bar\nu_e$, where $\ell=e,~\mu$ or $\tau$ from $W^+ \to \ell^+ \nu_\ell$.

Since only one top quark is produced, 
these final states have one $b$-jet so that they have
a negligible irreducible background, as mentioned previously.
There is, as before, a potentially dangerous reducible SM background
due to a reduced
$b$-tagging efficiency $ \epsilon_b $. For example, such a background
is generated by the reaction
$e^+e^- \to Z h $ when $h$ decays into a $ t
\bar t $ pair (assuming $ m_h > 2 m_t $) and one $b$ quark
in the top decay products is not detected. Similarly, for
$e^+e^- \to t \bar c \nu_e \bar\nu_e$ the SM reducible background
is generated by processes such as
$e^+e^- \to W^+W^- \nu_e \bar\nu_e,~t \bar t \nu_e \bar\nu_e$ 
(see also \cite{ttocmhdm3} and Hou {\it et al.} in \cite{tcprodmhdm}). 
Recall, however, that 
as in the case of $e^+e^- \to t \bar c$,
a $t \bar c$ signal have more experimental handles such 
as top reconstruction, a very energetic charm-jet, etc.
and these can be used to eliminate most of the background
events.

To obtain limits on the scale of the new 
physics, $\Lambda$, we 
again define the background-free cross sections, 
$\bar\sigma_{Ztc}$ and $\bar\sigma_{tc \nu \nu}$, 
by folding the $b$-tagging and top reconstruction
efficiency factors, which essentially eliminate 
the type of reducible backgrounds mentioned above.
Thus, our background-free observable cross sections are
\begin{eqnarray}
\bar\sigma_{tc \nu \nu} = \frac{2}{3} \epsilon_b \epsilon_t 
\sigma_{tc \nu \nu} ~~,~~
\bar\sigma_{Ztc} = \epsilon_b \epsilon_t \sigma_{Ztc} \label{htcbar1}~,
\end{eqnarray}
where the factor of 2/3 in the cross section for the 
$t \bar c \nu_e \bar\nu_e$ reaction takes into account the fact that
only the hadronic
top decay $t \to b j_1 j_2$ are useful (we assume that 
the semi-leptonic top decays 
cannot be reconstructed due to the additional two missing neutrinos 
in the final state).
As before, we assume that the largest value of $ \Lambda $ which can be
probed using these processes corresponds to the value yielding
a signal of 10 fully reconstructed events.

In Table 3 we give the $3\sigma$ 
limits that can be placed on the scale of the new 
physics $\Lambda$ using the processes $e^+e^- \to Z t \bar c + Z \bar t c$ 
and $e^+e^- \to t \bar c \nu_e \bar\nu_e + \bar t c \nu_e \bar\nu_e$, 
assuming no signal is observed (based on our 10 event criterion);
we take $\epsilon_b=60\%$, $\epsilon_t=80\%$ and 
$ a_L^h= a_R^h=1$.
We consider three collider scenarios: a NLC with $\sqrt s=500 $ GeV and 
a yearly integrated luminosity of $L=50$ $fb$$^{-1}$, $\sqrt s=1000 $ GeV with
$L=200$ $fb$$^{-1}$ and $\sqrt s=1500 $ GeV with $L=500$ $fb$$^{-1}$.
Entries marked by an $X$ in Table 3 indicate the cases 
for which no interesting limit can be obtained, i.e., where the limit 
corresponds to $\Lambda < \sqrt s$. 
Due to its decreasing nature, the 
cross section $e^+e^- \to Z t \bar c + Z \bar t c$ is only useful 
at 500 GeV, for which a limit of e.g., $\Lambda \gsim 830$ GeV is 
obtainable if $m_h \sim 250$ GeV. Using 
the $t \bar c \nu_e \bar\nu_e + \bar t c \nu_e \bar\nu_e$ 
final state, one can place the limits $\Lambda \gsim 1460$ GeV and 
$\Lambda \gsim 2140$ GeV in a 1000 GeV and 1500 GeV NLC, respectively (with
$m_h \sim 250$ GeV). Note that these limits are
weakened if $m_h=200$ or 400 GeV, since these
cross sections are smaller for such Higgs mass values 
(see Fig.~\ref{htcfig1}).\\

$$
\begin{tabular}{||c|c|c|c|c||} 
\hline
\multicolumn{5}{||c||}{Limits from $\{ \bar\sigma_{tc \nu \nu}~,~\bar\sigma_{Ztc}\}$}\\ \hline & 
&
$m_h=200$ GeV&
{$m_h=250$ GeV}&
{$m_h=400$ GeV}\\ 
$\sqrt{s}$ & 
$L$&
{}&
{}&
{}\\ 
\hline
500~GeV&
50~$fb^{-1}$&
$\{ 650~,~750\}$~GeV&
$\{ 650~,~830 \}$~GeV&
$\{X~,~X\}$~GeV
\\ \hline
1000~GeV&
200~$fb^{-1}$&
$\{1340~,~X \}$~GeV&
$\{1460~,~X \}$~GeV&
$\{1010~,~X \}$~GeV
\\ \hline
1500~GeV&
500~$fb^{-1}$&
$\{1930~,~X \}$~GeV&
$\{2140~,~X \}$~GeV&
$\{1600~,~X \}$~GeV
\\ 
\hline
\end{tabular}
$$

\bigskip
\bigskip

{\bf Table 3:} {\emph {Limits on the scale, $\Lambda$, of the new physics 
that generates new $htc$ effective operators, using 
the reactions $e^+e^- \to t \bar c \nu_e \bar\nu_e + \bar t c \nu_e \bar\nu_e$
and 
$e^+e^- \to Z t \bar c + Z \bar t c$ (in parenthesis).
The limits are given for $m_h=200,~250$ and $400$ GeV where 
in each case three accelerator scenarios are considered;
$\sqrt{s}=500$, $1000$ and $1500$ GeV with luminosities
$L=50$, $200$ and $500~fb^{-1}$, respectively. 
The signals considered are based on the total cross sections, 
as defined in (\ref{htcbar1}), 
assuming a $b$-tagging efficiency of $60\%$ and a top reconstruction 
efficiency of $80\%$. 
The limits are based on our criterion of $10$ events 
for the given luminosity and the given reaction (see also text).}}

\section{Right-handed \boldmath{$Wtb$} effects in \boldmath{$W^+W^- \to t \bar c$}}

The $WW$-fusion process $W^+W^- \to t \bar c$ can proceed at the 
tree-graph level in the SM via diagram (c) in Fig.~\ref{fig3}. The cross section,
however, is unobservably small due to GIM suppression:
$\sigma_{tc \nu \nu} \sim {\rm few} \times 10^{-4}$ $fb$ at a NLC
with c.m. energies in the range 1 - 2 TeV (see also \cite{ttocmhdm3}).

This suppression opens the possibility of 
observing a $t \bar c \nu_e \bar\nu_e$ signal in the presence of an effective 
right-handed $Wtb$ coupling, $\delta^t_{R,b}$, defined in (\ref{wdef}).
We consider these effects on 
the reaction $e^+e^- \to W^+W^- \nu_e \bar\nu_e \to 
 t \bar c \nu_e \bar\nu_e + \bar t c \nu_e \bar\nu_e$, 
which we evaluate using the EVBA.

We find, however, that the effective interactions do not
produce a significant enhancement in these cross sections
since $\delta^t_{R,b} \lsim 1$ (resulting from  $\alpha_{\phi \phi}$). 
The reason follows from the structure of 
the amplitude, ${\cal M}_{Wtd_i}$, 
for $W^+W^- \to t \bar c$ calculated using (\ref{wdef})
\begin{eqnarray} 
{\cal M}_{Wtd_i} &\propto& m_t \left( {\cal C}_{LL} (V_{ti} + \delta_{L,i}^t) 
(V_{ci}^* + \delta_{L,i}^c) + {\cal C}_{RR} \delta_{R,i}^t \delta_{R,i}^c \right)
\nonumber\\ 
&&+ m_{d_i} \left( {\cal C}_{LR} (V_{ti} + \delta_{L,i}^t) \delta_{R,i}^c +
{\cal C}_{RL} \delta_{R,i}^t (V_{ci}^* + \delta_{L,i}^c) \right) ~,
\end{eqnarray}
where ${\cal C}_{LL},~{\cal C}_{RR},~{\cal C}_{LR}$ and ${\cal C}_{RL}$ are some kinematic
functions with a mass dimension $-1$. 
If the only non-vanishing effective coupling is $\delta^t_{R,b}$, 
then the amplitude is proportional to the 
very small SM off-diagonal CKM element $V_{cb}$ and, in addition, it
contains a mass insertion factor $m_b$ 
from the t-channel $b$-quark propagator (see Fig.~\ref{fig3}(c)). 
If in addition $ \delta^c_{R,b}\not=0 $, then  
the amplitude receives also a contribution
proportional to $\delta^t_{R,b} \times \delta^c_{R,b}$ 
(with no mass insertion).
However, such a term will give a cross section which is proportional to 
$v^8/\Lambda^8$ instead of $v^4/\Lambda^4$ and is, therefore, also 
very small. 
We conclude that such right-handed current effects cannot 
be probed via the $WW$-fusion process. 

Before summarizing we wish to note that
the hard cross section $ W^+W^-  \to 
 t \bar c $ needed in the EVBA, exhibits a {\em physical} 
$t$-channel singularity \cite{tsingulr}. 
Due to the specific kinematics of this $2 \to 2$ process, 
the square of the $t$-channel momentum can be positive and the down quark 
propagator can, therefore, resonate once $t \sim m_d^2$. 
The reason for that is rather clear:
the incoming $W$-boson can decay to an on-shell 
pair of $d_i \bar c$ ($d_i=d,~s$ or $b$). The singularity, therefore, 
signals the production of an on-shell down quark in the $t$-channel.

The $t$-channel singularity of the $2 \to 2$ 
sub-process does not occur in the full 
$2 \to 4$ process. In the exact calculation, i.e., 
without using the EVBA, the exchanged 
$W^+$ and $W^-$ cannot be on-shell
since the $W^+,~W^-$ momenta are 
always space-like; as a consequence
the $Q^2$ of the $t$-channel down quark is always negative.\footnote{We
thank David Atwood for his helpful remarks regarding this point.}
Therefore, the EVBA, which assumes on-shell incoming vector-bosons, 
breaks down in such situations and cannot be used to approximate these
type of processes. 
To bypass this problem, we have used the EVBA with massless incoming 
$W$-bosons when calculating the above cross sections. 
We have checked that such an additional approximation gives rise to an error 
of the order of $\sim m_W/\sqrt s$ which is less than 
$10\%$ for a c.m. energy of $\sqrt s = 1000$ GeV. 

\section{Summary}

We have considered production of a $t \bar c$ pair in $e^+e^-$ 
colliders in the effective Lagrangian description.
We investigated a variety of processes, leading to a $t \bar c$ 
signal, which may be driven 
by some underlying flavor physics beyond the SM that gives rise to new
 vertices such as $Ztc$, $htc$, right-handed $Wtb$ and four-Fermi $tcee$
interactions. 

We have shown that, if present, 
the contributions of four-Fermi operators strongly
dominate the cross section for the 
reaction $e^+e^- \to t \bar c$, while the effects of flavor-changing $Z$
vertices are subdominant, assuming both types of effective operators
appear with coefficients of order one (which is the case in
all natural theories) and of similar scales though, as was mentioned 
previously, these two types of vertices may probe different kinds of physics 
and, therefore, should be measured separately.  

Thus, the $Ztc$ vertex may alternatively be
probed via the $t$-channel $WW$-fusion process $W^+W^- \to t \bar c$ 
which may yield an observable $t \bar c \nu_e \bar\nu_e$ signal at $1.5-2$ 
TeV $e^+e^-$ linear colliders. At 
hadron colliders the $Ztc$ vertex can also 
be efficiently probed  in flavor changing top decays 
\cite{han}, and in single top production in association with 
a $Z$-boson \cite{9906462}.

The $t$-channel $WW$-fusion process 
was also found to be sensitive to new $htc$ scalar interactions 
which may also lead to a $t \bar c \nu_e \bar\nu_e$ signal
at $1-2$ TeV NLC. We showed, however, that, at c.m. energies below 1 TeV, 
effective $htc$ couplings are 
better probed via the Bjorken process $e^+e^- \to Zh$ followed by
$h \to t \bar c$.

The effects of a new right-handed $Wtb$ coupling were found to be negligible 
for $t \bar c$ production in $e^+e^-$ colliders in $WW$-fusion 
processes. 

We have argued that, due to its unique characteristics,
the $t \bar c$ final state is essentially free of SM irreducible 
background and may be, therefore, easily identified in an 
$e^+e^-$ collider environment. 
In addition, by tagging the single $b$-jet coming from $t \to bW$ and 
by reconstructing the top quark from its decay products one is 
able, in principle, to eliminate 
all possible SM reducible background to the $t \bar c$ signal.
 
Using reasonable $b$-jet tagging and top reconstruction efficiencies 
at $e^+e^-$ colliders, we have derived sensitivity limits for
these machines to 
the scale of new flavor-changing physics, $\Lambda$.
For example, we find that an absence of a
$e^+e^- \to t \bar c$ signal at the recent 189 GeV LEP2 run already 
places the limit of $\Lambda \gsim 0.7 ~ (1.4)$ TeV on vector-like
(tensor-like) four-Fermi effective operators.
Similarly, the future 
200 GeV LEP2 run can place a limit 
of $\Lambda \gsim 1.5 ~ (2.5) $ TeV 
and, at a 1000 GeV NLC, the corresponding limits
are remarkably strong: $\Lambda \gsim 17 ~ (27)$ TeV; better
(due to a negligible SM background)
than those obtainable for flavor diagonal four-Fermi operators,
such as $ttee$.

Finally, concerning the limits on the scale of the effective 
operators that give rise to new 
$htc$ scalar interaction, we found, for example, 
that $\Lambda \gsim 830$ GeV at a 500 GeV NLC via the Bjorken process, and 
$\Lambda \gsim 2150$ GeV at a 1.5 TeV NLC via the $WW$-fusion process, 
if the mass of the SM Higgs is $\sim 250$ GeV. 

\begin{center}
{\bf Acknowledgments}
\end{center}

We thank D. Atwood, G. Eilam and A. Soni for discussions.
This research was supported in part by US DOE contract 
number DE-FG03-94ER40837(UCR).



\begin{thebibliography}{99}

\bibitem{nlcreviews} 
For recent reviews on linear colliders see:
ECFA/DESY LC Physics Working Group, E. Accomando {\it et al.}, 
Phys.\ Rep.\ {\bf 299}, 1 (1998); 
NLC ZDR Design Group and NLC Physics Working Group, S. Kuhlman {\it et al.}, 
hep-ex/9605011; 
H. Murayama and M. E. Peskin, Ann.\ Rev.\ Nucl.\ Part.\ Sci.\ 
{\bf 49}, 513 (1996).

\bibitem{ttocsm1} G. Eilam, J.L. Hewett and A. Soni, 
Phys.\ Rev.\ {\bf D44}, 1473 (1991); {\bf D59}, 039901(E) (1998).

\bibitem{ttocsm2} 
W. Buchm\"{u}ller and M. Gronau, 
Phys.\ Lett.\ {\bf 220B}, 641 (1989);
H. Fritzsch, Phys.\ Lett.\ {\bf 224B}, 423 (1989);
J.L. Diaz-Cruz, R. Martinez, M.A. Perez and 
A. Rosado, Phys.\ Rev.\ {\bf D41}, 891 (1990); 
B. Dutta-Roy {\it et al.}, 
Phys.\ Rev.\ Lett.\ {\bf 65}, 827 (1990); 
J.L. Diaz-Cruz and G. Lopez Castro, Phys.\ Lett.\ {\bf 301B}, 405 (1993);
B. Mele, S. Petrarca and and A. Soddu, Phys.\ Lett.\ {\bf 435B}, 401 (1998).

\bibitem{tcprodsm} 
A. Axelrod, Nucl.\ Phys.\ {\bf B209}, 349 (1982);
M. Clements {\it et al.}, Phys.\ Rev.\ {\bf D27}, 570 (1983);
V. Ganapathi {\it et al.}, Phys.\ Rev.\ {\bf D27}, 579 (1983);
G. Eilam, Phys.\ Rev.\ {\bf D28}, 1202 (1983);
C.-H. Chang {\it et al.}, Phys.\ Lett.\ {\bf 313B}, 389 (1993);
C.-S. Huang, X.-H. Wu and S.-H. Zhu, 
Phys.\ Lett.\ {\bf 452B}, 143 (1999).

\bibitem{ttocmhdm1} 
W.-S. Hou, Phys.\ Lett.\ {\bf 296B}, 179 (1992); 
M. Luke and M.J. Savage, 
Phys.\ Lett.\ {\bf 307B}, 387 (1993); 
J.L. Diaz-Cruz {\it et al.}, hep-ph/9903299.

\bibitem{ttocmhdm2} D. Atwood, L. Reina and A. Soni,
Phys.\ Rev.\ {\bf D55}, 3156 (1997).

\bibitem{ttocmhdm3} S. Bar-Shalom, G. Eilam, A. Soni and J. Wudka, 
Phys.\ Rev.\ Lett.\ {\bf 79}, 1217 (1997); {\it ibid.} 
Phys.\ Rev.\ {\bf D57}, 2957 (1998).

\bibitem{ttocsusy} M.J. Duncan, 
C.S. Li, R.J. Oakes and J.M. Yang, Phys.\ Rev.\ {\bf D31}, 1139 (1985);
Phys.\ Rev.\ {\bf D49}, 293 (1994), {\bf D56}, 3156(E) (1997);
J.-M. Yang and C.-S. Li, 
Phys.\ Rev.\ {\bf D49}, 3412 (1994), {\bf D51}, 3974(E) (1995). 
G. Couture, C. Hamzaoui and H. K\"{o}nig, 
Phys.\ Rev.\ {\bf D52}, 1713 (1995);
J.L. Lopez, D.V. Nanopoulos and R. Rangarajan, 
Phys.\ Rev.\ {\bf D56}, 3100 (1997);
G. Couture, M. Frank and H. K\"{o}nig, 
Phys.\ Rev.\ {\bf D56}, 4213 (1997);
G.M. de Divitiis, R. Petronzio and L. Silvestrini, 
Nucl.\ Phys.\ {\bf B504}, 45 (1997).

\bibitem{ttocrp1} J.M. Yang, B.-L. Young and X. Zhang, 
Phys.\ Rev.\ {\bf D58}, 055001 (1998).

\bibitem{ttocrp2} S. Bar-Shalom, G. Eilam and A. Soni,
to appear in Phys.\ Rev.\ {\bf D59} (1999). 

\bibitem{tcprodmhdm} D. Atwood, L. Reina and A. Soni, Phys.\ Rev.\ Lett.\ 
{\bf 75}, 3800 (1995); {\it ibid.} Phys.\ Rev.\ {\bf D53}, 1199 (1996); 
W.-S. Hou and G.-L. Lin, Phys.\ Lett.\ {\bf 379B}, 261 (1996);
J. Yi {\it et al.}, Phys.\ Rev.\ {\bf D57}, 4343 (1998). 
W.-S. Hou, G.-L. Lin and C.-Y. Ma, 
Phys.\ Rev.\ {\bf D56}, 7434 (1997); M. Sher, hep-ph/9809590. 

\bibitem{tcprodrp}
Z.-H. Yu {\it et al.}, hep-ph/9903471;
M. Chemtob and G. Moreau, Phys.\ Rev.\ {\bf D59}, 116012 (1999); 
U. Mahanta and A. Ghosal, 
Phys.\ Rev.\ {\bf D57}, 1735 (1998). 

\bibitem{prl82p1628} F. del Aguila, 
J. A. Aguilar-Saavedra and R. Miquel,
Phys.\ Rev.\ Lett.\ {\bf 82}, 1628 (1999).

\bibitem{b.and.w} 
W.~Buchmuller and D.~Wyler, Nucl. Phys. {\bf B268}, 621 (1986);
C. Arzt, M.B. Einhorn and J. Wudka, Nucl.\ Phys.\ {\bf B433}, 41 (1995).

\bibitem{bparity} S. Bar-Shalom and J. Wudka, hep-ph/9904365.

\bibitem{eetcindep1} K. Hikasa, Phys.\ Lett.\ {\bf 149B}, 221 (1984).

\bibitem{joan} T. Han and J.L. Hewett, hep-ph/9811237.

\bibitem{hep9712394} V.F. Obraztsov, S.R. Slabospitsky and 
O.P. Yushchenko, Phys.\ Lett.\ {\bf 426B}, 393 (1998).

\bibitem{davidsher} D. Atwood and M. Sher, 
Phys.\ Lett.\ {\bf 411B}, 221 (1997).

\bibitem{gamgamtc} K.J. Abraham, K. Whisnant and B.-L. Young,
Phys.\ Lett.\ {\bf 419B}, 381 (1998). 

\bibitem{jose2} A. Djouadi {\it et al.}, in 
{\it Physics and Experiments with Linear Colliders}, eds. R. Orava 
{\it et al.}, Saariseka, Finland,
9-14 Sept. 1991 (World Scientific, Singapore, 1992).
Int. J. Mod. Phys. {\bf A} (Proc. Suppl.) {\bf1A \&\ B}, 1993.
 

\bibitem{grzad} B. Grzadkowski, Acta.\ Phys.\ Polon.\ {\bf B27}, 921 (1996);
B. Grzadkowski, Z. Hioki and M. Szafranski, 
Phys.\ Rev.\ {\bf D58}, 035002 (1998).

\bibitem{naturality} A. Manohar and H. Georgi,
Nucl.\ Phys.\ {\bf B234}, 189 (1984).

\bibitem{evba} S. Dawson, Nucl.\ Phys.\ {\bf B249}, 42 (1985);
S. Dawson and S. S. D. Willenbrock, Nucl.\
Phys.\ {\bf B284}, 449 (1987); 
P.W. Johnson, F.I. Olness and W.-K. Tung, Phys.\ Rev.\ {\bf D36}, 291 (1987);
R.P. Kauffman, Phys.\ Rev.\ {\bf D41}, 3343 (1990).

\bibitem{collider} See e.g., V. Barger and R. Phillips,
{\it Collider Physics}, (Addison-Wesley;
Redwood City, Calif.; 1987).

\bibitem{frey} See e.g., R. Frey, hep-ph/9606201, in proceedings
of the workshop on Physics and Experiments with Linear Colliders, edited by
A. Miyamoto, Y. Fujii, T. Matsui and S. Iwata 
(World Scientific, Singapore, 1996).

\bibitem{shen} B. Shen, private communication. 

\bibitem{peskin} C.R. Schmidt and M. Peskin, Phys.\ Rev.\
Lett.\ {\bf69}, (1992) 410.

\bibitem{topspinhad} B. Grzadkowski and J.F. Gunion, 
Phys.\ Lett.\ {\bf B350}, (1995) 218.

\bibitem{pdg} C. Caso et al., Eur. Phys. J. {\bf C3}, 1 (1998). 

\bibitem{higgshunters} See e.g., 
J.F. Gunion, H.E. Haber, G. Kane and S. Dawson,
{\it The Higgs Hunter's Guide} (Addison-Wesley;
Redwood City, Calif.; 1990). 

\bibitem{tsingulr} For examples of other types of $t$-channel singularities 
see  
R.F. Pierls, Phys.\ Rev.\ Lett.\ {\bf6}, 641 (1961);
I.F.~Ginzburg, hep-ph/9509314;
K.~Melnikov and V.G.~Serbo, Nucl. Phys. {\bf B483}, 67 (1997).

\bibitem{han} T. Han, R.D. Peccei and X. Zhang,
Nucl. Phys. {\bf B454}, 527 (1995).

\bibitem{9906462} F. del Aguila, J. A. Aguilar-Saavedra and 
Ll. Ametller, hep-ph/9906462.

\end{thebibliography}
\end{document}